\documentclass[
 reprint,
 superscriptaddress,
 amsmath,amssymb,
 prx,
 pra,
]{revtex4-2}

\usepackage[english]{babel}
\usepackage{graphicx}
\usepackage{dcolumn}
\usepackage{bm}

\usepackage[utf8]{inputenc}
\usepackage[T1]{fontenc}

\usepackage{braket}
\usepackage[caption=false]{subfig}
\usepackage{booktabs}
\usepackage{amsmath}
\usepackage{cancel}

\usepackage{comment}

\begin{document}

\title[Admittance extraction of quantum dot arrays for charge sensing]{Admittance extraction of quantum dot arrays for charge sensing}
\author{Conor Power}
\email{conor.power@equal1.com}
\affiliation{Equal1 Laboratories}
\affiliation{School of Electronic and Electrical Engineering, University College Dublin}%

\author{Mathieu Moras}
\affiliation{Equal1 Laboratories}

\author{Andrii Sokolov}
\affiliation{Equal1 Laboratories}

\author{Robert Bogdan Staszewski}
\affiliation{Equal1 Laboratories}
\affiliation{School of Electronic and Electrical Engineering, University College Dublin}%

\author{Elena Blokhina}
\affiliation{Equal1 Laboratories}
\affiliation{School of Electronic and Electrical Engineering, University College Dublin}%

\date{\today}%

\begin{abstract}
This work demonstrates admittance extraction of quantum dot arrays using a rate equation framework matching experimental reflectometry measurements. Using a semiclassical rate equation, we consider a perturbation of the source/drain chemical potential to model reflectometry readout in a quantum dot array. This is a time-dependent extension of the rate equation which has been employed in literature for pumped arrays of quantum dots and capacitance extractions. Here we extend this to a double quantum dot for admittance extraction. The simulations are calibrated using Fermi-Hubbard parameters extracted from experimental data, which it is then shown to closely approximate to. 
\end{abstract}

\maketitle

\section{Introduction}

Quantum dot arrays in commercial semiconductor processes are a promising approach to scalable implementations of spin qubits. The measurement of quantum dot-based qubits is typically done through sensing electrometers capacitively coupled to the quantum dot structure or through the sensing of the quantum dot structure directly. Sensing in this context refers to either the monitoring of a baseband voltage/current\cite{power_charge-based_2024} or the reflection coefficient using reflectometry methods\,\cite{vigneau_probing_2022}. A number of these devices are also used for temperature sensing in literature\,\cite{frake_radio-frequency_2015,ahmed_primary_2018, cochrane_intrinsic_2024}. Reflectometry methods are typically much faster than baseband methods in existing experimental setups, however reflectometry requires embedding the quantum dots in an LC cavity which can occupy a large footprint. However, there have been recent efforts to reduce this footprint and move the inductor from off-chip to on-chip\,\cite{swift_superinductor_2025}. Therefore, as we continue to scale the number of quantum dots we must also have fast methods to assess and improve measurement methods. There are also efforts towards the co-simulation of the reflectometry response of quantum dot arrays alongside classical electronic circuits using the Lindblad equation\,\cite{peri_unified_2024, peri_compact_2025}.

In this work we propose a rate equation model of reflectometry, allowing for the prediction of various electrometer responses, such as single-electron transistors and single-electron boxes. The single-electron box devices are of interest as they only require a single reservoir and reduce the electrometer footprint. We consider reflectometry performed on the reservoirs in this work so as to align with our experimental data. The experimental data presented in this work comes from fully-tunable quantum dot arrays in the commercial 22FDX process from GlobalFoundries. We first demonstrate the calibration of the model against known steady-state results of common charge sensing architectures, followed then by an extension to a time-varying perturbation to capture the response of reflectometry. There is close alignment with experimental data for both the Double-Island Single-Electron Transistor and the Double-Island Single-Electron Box, along with an experimental prediction for future demonstration of near-equilibrium Pauli-Spin Blockade in 22FDX.

\section{Hamiltonian Model}
We will first define the most general Hamiltonian that we use in this work. $\hat{H}_{S}$ describes two tunnel-coupled electron-spin orbitals with Coulomb and Zeeman interactions, $\hat{H}_{I}$ then describes the interaction between each lead and the adjacent spin orbitals, and $\hat{H}_{B}$ describes the leads themselves with a probe perturbation due to reflectometry. The total Hamiltonian is the sum of these:

\begin{equation}\label{eqn:refleto_total_ham}
    \hat{H} = \hat{H}_{\text{S}} + \hat{H}_{\text{B}} + \hat{H}_{\text{I}}
\end{equation}
where each of these terms are defined as follows:

\begin{widetext}
\begin{align}
    \hat{H}_{\text{B}} =& \sum_{\substack{k\sigma\\b=S,D}}(\underbrace{\epsilon_{k\sigma b} - eV_{ac}^{b}\sin(\omega t)}_{\text{Lead + Probe Perturbation}})\hat{c}_{k\sigma b}^{\dagger}\hat{c}_{k\sigma b}\\
    \hat{H}_{\text{I}} =& \sum_{k\sigma}(\underbrace{\tau_{k\sigma S}\hat{d}_{1\sigma}\hat{c}_{k\sigma S}^{\dagger} + \tau_{k\sigma S}^{*}\hat{c}_{k\sigma S}\hat{d}_{1\sigma}^{\dagger}}_{\text{Source-Dot 1 Interaction}})
                       + \sum_{k\sigma}(\underbrace{\tau_{k\sigma D}\hat{d}_{2\sigma}\hat{c}_{k\sigma D}^{\dagger} + \tau_{k\sigma D}^{*}\hat{c}_{k\sigma D}\hat{d}_{2\sigma}^{\dagger}}_{\text{Drain-Dot 2 Interaction}})\\
    \hat{H}_{\text{S}} = &\sum_{\substack{i=1,2\\\sigma=\uparrow,\downarrow\\b=S,D}}(\underbrace{\epsilon_{i}}_{\text{Onsite Potential}} -\underbrace{e\alpha_{i}^{b}V_{ac}^{b}\sin(\omega t)}_{\text{Reduced Onsite Probe}} -\underbrace{\sigma[g\mu_{B}B/2]}_{\text{Zeeman Splitting}})\hat{n}_{i\sigma} \nonumber\\+ &\sum_{i=1,2}\underbrace{\frac{U_{i}}{2}\hat{n}_{i}(\hat{n}_{i}-1)}_{\text{Onsite Coulomb Interaction}} + \underbrace{U_{12}\hat{n}_{1}\hat{n}_{2}}_{\text{Inter-Site Coulomb Interaction}}
    - \sum_{\sigma=\uparrow,\downarrow}\underbrace{t(\hat{d}_{1\sigma}^{\dagger}\hat{d}_{2\sigma} + \hat{d}_{2\sigma}^{\dagger}\hat{d}_{1\sigma})}_{\text{Inter-Site Tunnel Coupling}}
\end{align}
\end{widetext}

This Hamiltonian is expressed in occupation operator representation\,\cite{bruus_many-body_2016} and all operators are summarised in Tab.\,\ref{tab:operators}. Not all terms in the Hamiltonian will be applicable for all of the charge sensing architectures considered in this work. For example, when considering a single quantum dot, the inter-site Coulomb interaction and inter-site tunnel coupling terms will be zero. As presented, this Hamiltonian represents a double quantum dot with electron lead connections to each dot and a maximum of two electrons with spin per dot. We have included the probe perturbation term in the lead states to capture the effect of reflectometry on the system. We have also included a reduced onsite probe term in $\hat{H}_{S}$ to model the finite leverarm from the leads to the quantum dots due to the probe perturbation. The size of the perturbation is chosen close to experimental values and also calibrated by inspecting when the amplitude no longer results in a linear response, as discussed in Appendix.\,\ref{app:linearresponse}.

\begin{table}[b]
    \caption{Summary of creation, annihilation, and number operators in Hamiltonian \eqref{eqn:refleto_total_ham}. The subscripts index over the following; $k$ over the momentum of the infinite non-interacting electron reservoirs, $i \in \{1,2\}$ over the quantum dots, $\sigma \in \{\uparrow,\downarrow\}$ over the spin state, and $b \in \{S,D\}$ over the two non-interacting reservoirs.}
    \label{tab:operators}
    \centering
    \footnotesize
    \setlength{\tabcolsep}{4pt}
    
    \raggedright
    \begin{tabular*}{\columnwidth}{@{}l@{\extracolsep{\fill}}l@{}}
    \toprule
    Operator & Physical Meaning \\
    \midrule
    $\hat c_{k\sigma b}$ & \parbox[t]{0.52\columnwidth}{\raggedright Annihilate electron with momentum $k$ and spin $\sigma$ in lead $b$} \\
    $\hat c^\dagger_{k\sigma b}$ & \parbox[t]{0.52\columnwidth}{\raggedright Create electron with momentum $k$ and spin $\sigma$ in lead $b$} \\
    $\hat d_{i\sigma}$ & \parbox[t]{0.52\columnwidth}{\raggedright Annihilate electron with spin $\sigma$ on dot $i \in \{1,2\}$} \\
    $\hat d^\dagger_{i\sigma}$ & \parbox[t]{0.52\columnwidth}{\raggedright Create electron with spin $\sigma$ on dot $i$} \\
    $\hat n_{i\sigma} = \hat d^\dagger_{i\sigma}\hat d_{i\sigma}$ & \parbox[t]{0.52\columnwidth}{\raggedright Number operator counting electrons of spin $\sigma$ on dot $i$} \\
    $\hat n_i = \sum_\sigma \hat n_{i\sigma}$ & \parbox[t]{0.52\columnwidth}{\raggedright Total electron number on dot $i$ (spin-summed)} \\
    \bottomrule
    \end{tabular*}
\end{table}

The mapping of occupation number operators to qubits is then done for finding the matrix representation of the creation and annihilation operators for numerical simulation. This is achieved using the Jordan-Wigner transform. Sequential tunnelling events in the semiclassical transport framework, introduced in the next section, allow us to traverse the many-body states one electron at a time. Note that $\hat{H}_{B}$ and $\hat{H}_{I}$ include sums over an infinite number of electrons in the reservoirs, so we approximate these next with a semiclassical rate equation. 

\section{Semiclassical Approximation}\label{sec:reflecto_semiclassicalapprox}

The key to approximating the Hamiltonian in Eq.\,\eqref{eqn:refleto_total_ham} is to perform perturbation theory in the lead-dot interaction, also known as the weak-coupling approximation. This is carried out in detail in existing literature\,\cite{ihn_semiconductor_2015, schaller_open_2014} so we won't repeat this derivation here. %
This results in the Pauli Master Equation (PME) or Quantum Optical Master Equation, which describes how the populations of the density matrix change over time in the weak coupling approximation\,\cite{campaioli_tutorial_2023}. The density matrix coherences go to zero in this approximation\,\cite{fransson_pauli_2006, petropoulos_nanoscale_2024}. Thermalisation then occurs through the leads and the steady state solution of the PME matches the thermal Boltzmann distribution over the system eigenvalues\,\cite{schaller_open_2014}. 

The infinite reservoirs of non-interacting electrons act as approximations for degenerately doped regions in semiconductors. Due to the interaction with these reservoirs in the weak coupling approximation, electrons can be exchanged between the dot region and the reservoirs. Intuitively, the PME can be written by considering the change in a single population, $p_{N\alpha}(t)$, corresponding to the population of $\ket{N,\alpha}$, the $\alpha$ many-body state within the $N$ electron subspace. The key here is that the many-body states are found through diagonalising only $\hat{H}_{\text{S}}$ from the total Hamiltonian. The effects of the lead-dot and lead Hamiltonians, $\hat{H}_{\text{I}}$ and $\hat{H}_{\text{B}}$, are accounted for in the tunnel rates we introduce now. The population $p_{N\alpha}(t)$  of the $\ket{N,\alpha}$ many-body state increases as other states transition from $\ket{L,\beta}$ to $\ket{N,\alpha}$ with rate $\Gamma_{N\alpha, L\beta}$ and decreases as the $\ket{N,\alpha}$ state transitions to $\ket{M,\gamma}$ with rate $\Gamma_{M\gamma,N\alpha}$. We denote the cardinality of the $N$ electron subspace as $|N|$ and the maximum number of electrons as $\#e$. 
\begin{align}
    \frac{d}{dt}p_{N\alpha}(t) &= \sum_{L=0}^{\#e}\sum_{\beta=0}^{|L|-1}\Gamma_{N\alpha,L\beta}(t)p_{L\beta}(t)\nonumber\\&\quad - \sum_{M=0}^{\#e}\sum_{\gamma=0}^{|M|-1}\Gamma_{M\gamma,N\alpha}(t)p_{N\alpha}(t)
\end{align}

The rates between populations can be assembled into a rate matrix. It is important to note that given that we are considering only transition rates between the populations of the density matrix, these are exactly the eigenstates of the closed-system Hamiltonian with the influence of the leads included via transition rates. The rate equation in general is only valid when working in the eigenbasis of the system\,\cite{schaller_open_2014}. The rate matrix form of the PME is then written as:
\begin{equation}\label{eqn:rate_eq_matrix}
    \frac{d}{dt}\textbf{p}(t) = \bm{\Gamma}(t)\textbf{p}(t)
\end{equation}
where $\textbf{p}(t)$ is the vector of populations and $\bm{\Gamma}$(t) is the rate matrix defined as follows:
\begin{gather}
    \textbf{p}(t) = \begin{pmatrix}p_{00}(t)\\p_{10}(t)\\\vdots\\p_{N0}(t)\end{pmatrix},
    \bm{\Gamma}(t) = \begin{pmatrix}
    -\Sigma_{00} & \Gamma_{00,10} & \cdots & \Gamma_{00,N0} \\
    \vdots & \vdots & & \vdots \\
    \Gamma_{N0,00} & \Gamma_{N0,10} & \cdots & -\Sigma_{N0}
    \end{pmatrix}
    \raisetag{-1.5\baselineskip}
\end{gather}
where we define the diagonal sum terms as $\Sigma_{00} = \sum_{M=0}^{\#e} \sum_{\gamma=0}^{|M|-1} \Gamma_{M\gamma,00}$ and $\Sigma_{N0} = \sum_{M=0}^{\#e} \sum_{\gamma=0}^{|M|-1} \Gamma_{M\gamma,N0}$.

The steady state solution to Eqn.\,\eqref{eqn:rate_eq_matrix} can be found by considering the time rate of change of the populations, $\textbf{p}(t)$, to be equal to zero. This then implies;
\begin{equation}\label{eqn:steady_state_sol}
    \bm{\Gamma}(t)\textbf{p}(t) = 0
\end{equation}
We use this equation to find the steady state current through a number of different quantum dot array setups in the supplementary material and as the initial condition for the reflectometry response.

There are 4 possible charge occupation basis states for a single quantum dot with spin (for a total of 2 orbitals). Then in the case of a double quantum dot with spin (or 4 orbitals), there are 16 possible charge occupation basis states, shown in Tab.\,\ref{tab:4-electron-subspaces}. We denote the 4 orbital state as $\ket{n_{\uparrow1},n_{\downarrow1},n_{\uparrow2},n_{\downarrow2}}$, where $n_{\uparrow1} \in \{0,1\}$ denotes the occupation of orbital 1 by a spin $\uparrow$ electron.

\begin{table}[t]
    \centering
    \begin{tabular}{c|l}
        \# electrons & Occupation Basis States \\
        \hline
        0 & $\lvert 0000 \rangle$ \\
        1 & $\lvert 0001 \rangle,\, \lvert 0010 \rangle,\, \lvert 0100 \rangle,\, \lvert 1000 \rangle$ \\
        2 & $\lvert 0011 \rangle,\, \lvert 1100 \rangle,\, \lvert 0101 \rangle,\, \lvert 1010 \rangle,\, \lvert 1001 \rangle,\, \lvert 0110 \rangle$ \\
        3 & $\lvert 0111 \rangle,\, \lvert 1110 \rangle,\, \lvert 1101 \rangle,\, \lvert 1011 \rangle$ \\
        4 & $\lvert 1111 \rangle$ \\
        \hline
    \end{tabular}
    \caption{All sixteen of the possible occupation number basis states. In general, eigenstates of the system Hamiltonian are superpositions of these states. }
    \label{tab:4-electron-subspaces}
\end{table}

We now write an individual rate between a bath of non-interacting electrons and a quantum dot using Fermi's Golden Rule as follows, where the sum is over the states specified in the lead-dot interaction term\,\cite{ihn_semiconductor_2015,bruus_many-body_2016}. The object, $\Gamma_{(N+1)\beta,N\alpha}^{S}$, describes the rate of transition between the $\alpha$ state of the $N-$electron subspace and the $\beta$ state of the $N+1$ electron subspace by the addition of a single electron via the source lead, S:
\begin{align}
    \Gamma^{S}_{(N+1)\beta,N\alpha} &= \gamma^S 
    \left| \bra{N+1,\beta} \sum_{\sigma=\uparrow,\downarrow} d_{1\sigma}^\dagger \ket{N,\alpha}\right|^2\nonumber\\&\quad\times
    n_F(E_{(N+1)\beta} - E_{N\alpha} - \mu_S)\label{eqn:source_rate_add}\\
    \Gamma^{S}_{(N-1)\beta,N\alpha} &= \gamma^S 
    \left| \bra{N-1,\beta} \sum_{\sigma=\uparrow,\downarrow} d_{1\sigma} \ket{N,\alpha}\right|^2 
    \nonumber\\&\quad\times\left(1 - n_F(E_{N\alpha} - E_{(N-1)\beta} - \mu_S)\right)\label{eqn:source_rate_rem}
\end{align}
We can also write similar rates due to interactions with the drain lead. However, in the case of the other lead, we sum only over the creation/annihilation operators of the dot adjacent to that lead. Notice that we only sum over the creation/annihilation operators for the dot adjacent to the source lead in Eqs.\,\eqref{eqn:source_rate_add} and \eqref{eqn:source_rate_rem}. The meanings of each new term in the rates are given in Tab.\,\ref{tab:tunnel_rates_explained}.

\begin{table}[h]
    \caption{Explanation of symbols appearing in the tunnelling rate expressions, \eqref{eqn:source_rate_add} and \eqref{eqn:source_rate_rem}.}
    \label{tab:tunnel_rates_explained}
    \centering
    \footnotesize
    \setlength{\tabcolsep}{4pt}
    
    \raggedright
    \begin{tabular*}{\columnwidth}{@{}l@{\extracolsep{\fill}}l@{}}
    \toprule
    \textbf{Symbol} & \textbf{Physical Meaning} \\
    \midrule
    $\gamma^{S}$ & \parbox[t]{0.65\columnwidth}{\raggedright Tunnel coupling strength between the system and lead $S$. This is a result of the wide band limit.}\\
    $\lvert N,\alpha\rangle$ & \parbox[t]{0.65\columnwidth}{\raggedright Many-body eigenstate $\alpha$ of the system with $N$ electrons.} \\
    $\bigl|\langle \cdots \rangle\bigr|^2$ & \parbox[t]{0.65\columnwidth}{\raggedright Squared many-body matrix element governing the tunnelling overlap.} \\
    $E_{N\alpha}$ & \parbox[t]{0.65\columnwidth}{\raggedright Energy eigenvalue of the many-body eigenstate $\lvert N,\alpha\rangle$.} \\
    $\mu_S$  & \parbox[t]{0.65\columnwidth}{\raggedright Chemical potential of lead $S$.} \\
    $n_F(x)$ & \parbox[t]{0.65\columnwidth}{\raggedright Fermi--Dirac distribution $n_F(x) = \left(e^{x/k_B T} + 1\right)^{-1}$ describing occupation of lead states.} \\
    \bottomrule
    \end{tabular*}
\end{table}

The total rate between a pair of many-body states is the sum over the contributions from all of the leads in the weak coupling regime\,\cite{schaller_open_2014}. Therefore, we state that the total tunnel rate between $\ket{N,\alpha}$ and $\ket{N+1,\beta}$ is the sum of the rates due to interactions with the source and the drain.
\begin{equation}
    \Gamma_{(N+1)\beta,N\alpha} = \Gamma_{(N+1)\beta,N\alpha}^{S} + \Gamma_{(N+1)\beta,N\alpha}^{D}
\end{equation}

We use the wide-band limit for the leads of non-interacting electrons for simplicity. The wide-band limit essentially treats the energy landscape of the leads as flat, in the sense that the hybridisation is taken to be constant for all lead states, $\tau_{S,\nu_S,\nu_d} = \tau$, and couples equally to all many-body states. In other words, the product of the hybridisation and the DOS, $g_S(\nu_{S})$, is considered constant\,\cite{bruus_many-body_2016}, resulting in $\gamma^{S}$:
\begin{equation}
    \gamma^{S} = \frac{2\pi}{\hbar}|\tau_{S,\nu_{S},\nu_{d}}|^{2}g_{S}(\nu_{S})
\end{equation}

Therefore, in the wide-band limit, we take a uniform hybridisation in the range of weak coupling\cite{petropoulos_nanoscale_2024}. 
We take a constant density of states at the Fermi level so we only need evaluate $g_{S}(E_{F})$. We use the simulation tool QTCAD\,\cite{noauthor_qtcad_nodate} to estimate this value. In a QTCAD simulation the Fermi level is taken as the zero reference so the energy difference between the Fermi level and the conduction band edge, $E_{C}$, is the quantity we extract from simulation. This difference gives an indication as to the availability of free charge carriers. So, from simulations in QTCAD, we see a typical $(E_{F} - E_{C}) \approx 0.02~\text{eV}$, using the calibration process described in\,\cite{sokolov_common-mode_2025}.

Given that we can now solve for the time-dependent populations using Eq.\,\ref{eqn:rate_eq_matrix} and the steady-state solution using Eq.\,\ref{eqn:steady_state_sol}, we note that we can also compute the current in this sequential tunnelling approximation. For all tunnelling events that increase the electron count in the dot region we consider this positive current, and for all tunnelling events that remove an electron from the dot region we consider this negative current. Note that electrons may leave and enter through any of the leads connected to the dot region. Here we will consider one or two leads connected to the dot region. The current is then defined in this rate equation approximation as follows:
\begin{align}
    I(t) &= \sum_{N=0}^{\#e}\left(\sum_{\beta = 0}^{|N+1|-1}\Gamma_{(N+1)\beta,N\alpha}(t)\right.\nonumber\\ &\quad\left.- \sum_{\gamma=0}^{|N-1|-1}\Gamma_{(N-1)\gamma,N\alpha}(t)\right)p_{N\alpha}(t)
\end{align}
where $\Gamma_{-1\gamma,0\alpha} = 0$ as we cannot have less than 0 electrons in the dot. Throughout the rest of this work we will focus on the four devices in Fig.\,\ref{fig:QuantumMeasurement_ReflectoDevices}. We show the steady state response in Appendix.\,\ref{sec:stationaryresults} and now move to extraction of admittance for reflectometry for the charge sensors. 

\begin{figure}[h]
    \centering
    \includegraphics[width=1.0\linewidth]{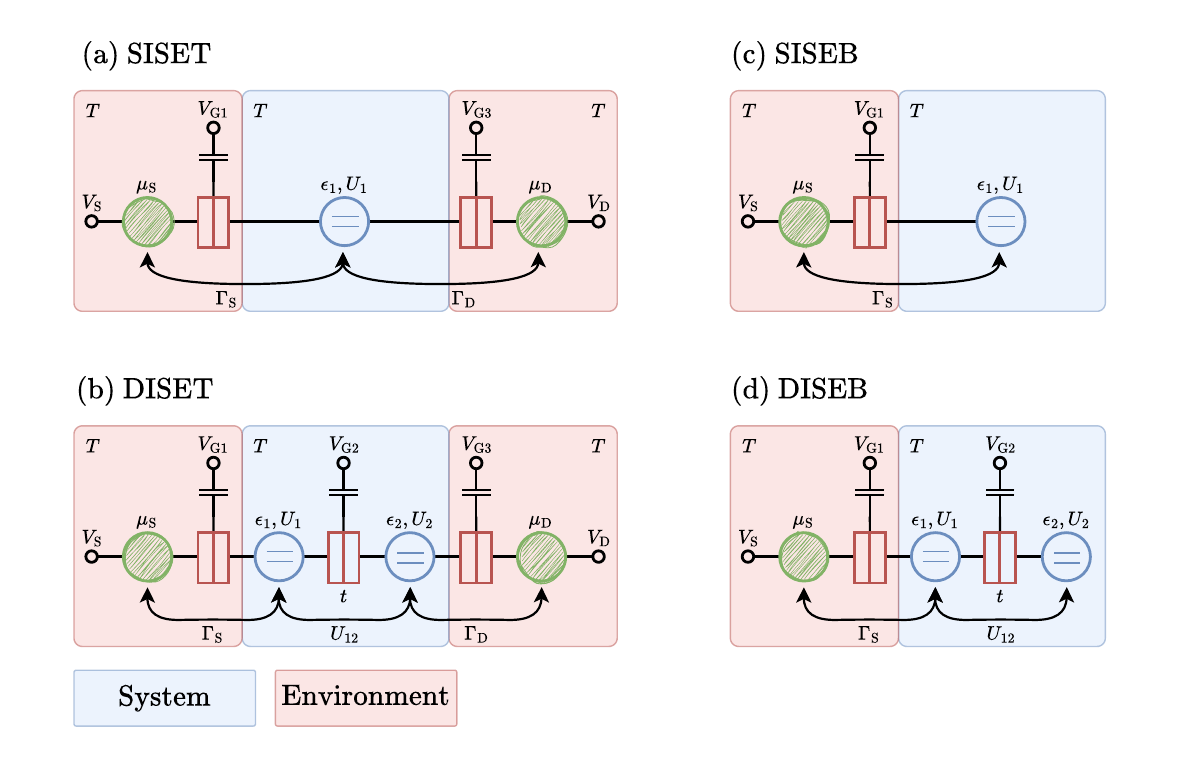}
    \caption{SET devices are those that allow current to flow from lead to lead, and SEB devices as those with only a single lead. (a) Single-Island Single-Electron Transistor (SISET)(b) Double-Island Single-Electron Transistor (DISET). (c) Single-Island Single-Electron Box (SISEB). (d) Double-Island Single-Electron Box (DISEB)}
    \label{fig:QuantumMeasurement_ReflectoDevices}
\end{figure}

\section{Reflectometry Modelling}
We now investigate rate equation modelling of reflectometry. Let's first understand how reflectometry works and how we then incorporate this into our model. Reflectometry is the term used in literature when measuring the reflection coefficient. We can probe our devices from many points such as the reservoirs or the gates and investigate the reflection coefficient. This gives us an insight into the impedance/admittance of the quantum dot structure under investigation. 

\begin{figure}[h]
    \centering
    \includegraphics[width=0.7\linewidth]{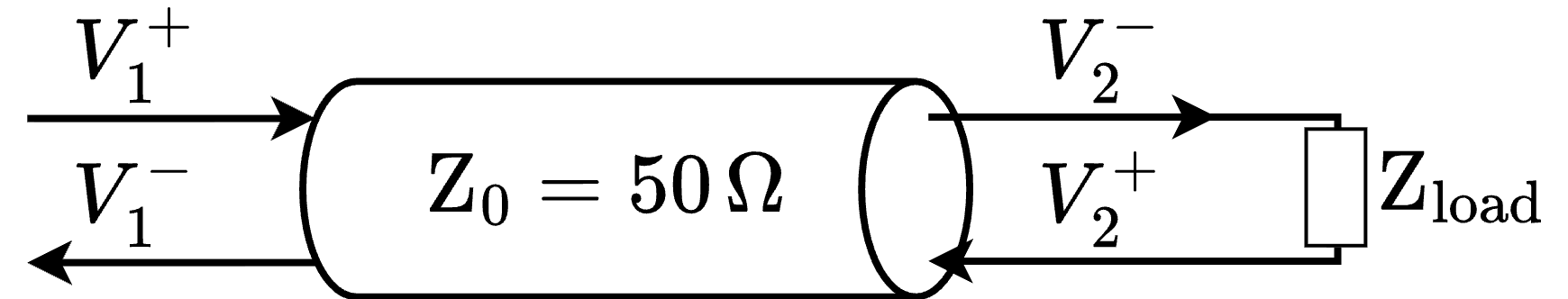}
    \caption{A coaxial cable with characteristic impedance $\text{Z}_{0}$ connected to a load impedance $\text{Z}_{\text{load}}$. Signals flowing into the cable are denoted with $+$ and those flowing out with $-$.}
    \label{fig:QuantumMeasurement_ReflectoCoaxCable}
\end{figure}

Reflectometry involves sending a time-varying voltage of a known frequency down a channel with a known characteristic impedance $\text{Z}_{0}$. We show the channel as a coaxial cable in Fig.\,\ref{fig:QuantumMeasurement_ReflectoCoaxCable}. This is then connected to a load impedance $\text{Z}_{\text{load}}$.

The reflection coefficient, $\Gamma$, not to be confused with the rates in Sec.\,\ref{sec:reflecto_semiclassicalapprox}, is the ratio of the reflected voltage waveform over the incident voltage waveform. This can be expressed in terms of the load impedance ($\text{Z}_{\text{load}}$) or admittance ($\text{Y}_{\text{load}} = 1/\text{Z}_{\text{load}}$) and the characteristic impedance of the line, $\text{Z}_{0}$:
\begin{equation}
    \Gamma = \frac{V_{1}^{-}}{V_{1}^{+}} = \frac{\text{Z}_{\text{load}} - \text{Z}_{0}}{\text{Z}_{\text{load}} + \text{Z}_{0}} = \frac{1 - \text{Z}_{0}\text{Y}_{\text{load}}}{1 + \text{Z}_{0}\text{Y}_{\text{load}}}
\end{equation}

An impedance $\text{Z}$ is in general complex and is written as $\text{R} + j\text{X}$ where $\text{R}$ is the resistance of the load and $\text{X}$ is the reactance of the load. We will typically work with admittance in simulation as in many scenarios the impedance goes to infinity, whilst the corresponding admittance goes to zero. This occurs since we are only modelling the quantum device, while in reality there are classical current contributions that would temper the extreme response of the quantum device. This is noted for investigation in future work.

\begin{figure}[h]
    \centering
    \includegraphics[width=1.0\linewidth]{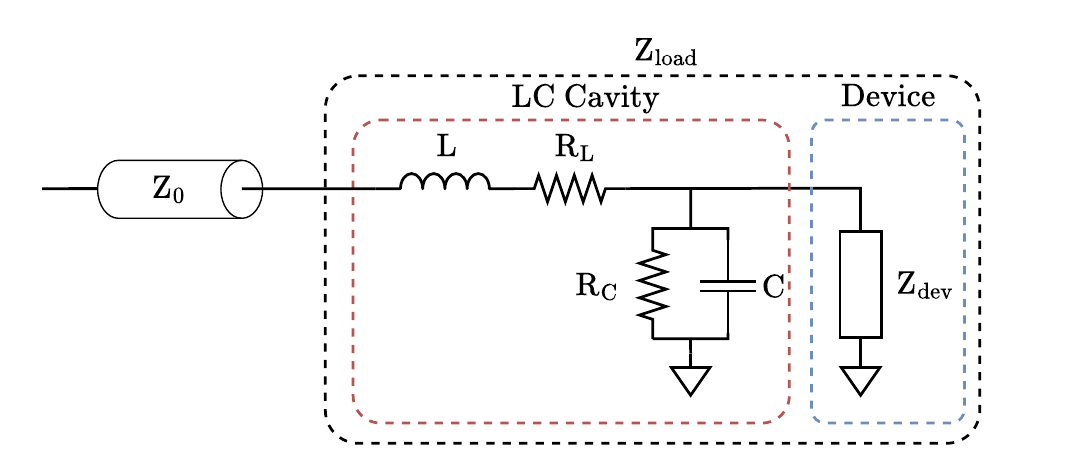}
    \caption{Circuit diagram of coaxial cable, LC cavity, and a device. $\text{Z}_{0}$ represents the coaxial cable. $\text{Z}_{\text{load}}$ is expanded into two parts; the LC cavity impedance and the device impedance, $\text{Z}_{\text{dev}}$.}
    \label{fig:QuantumMeasurement_ReflectoLCTankDevice}
\end{figure}

 In Fig.\,\ref{fig:QuantumMeasurement_ReflectoLCTankDevice} the load impedance consists of an LC cavity and a device we are interested in measuring. The LC tank is used to provide a known resonant response in the absence of the device impedance, $\text{Z}_{\text{dev}}$ and to match to the impedance of the coaxial line. Matching ensures the load impedance appears the same as the channel impedance for maximum power transfer\,\cite{vigneau_probing_2022}. In other words, the values of the components in the LC tank are chosen such that $\text{Z}_\text{load} \approx \text{Z}_{0}$. 

Depending on the nature of the device impedance (whether it is resistive, reactive, or both), the resonant response of the LC tank will change in different ways. This is the core principle behind reflectometry readout of charge sensors. In the rate equation framework we consider reflectometry as a time-dependent perturbation to the tunnel rates.

As we will see, it is possible to correlate changes in the reflection coefficient with the state of quantum dot devices. We investigate this for the DISET and DISEB in Fig.\,\ref{fig:QuantumMeasurement_ReflectoDevices}. All of these devices exhibit both resistive and reactive variations. This is to a greater or lesser degree depending on whether we consider a SET or SEB type device. In a SET device, we would expect resistive variations to be more prominent as current flows through the entire device. However, in a SEB device, current only flows through a single reservoir and is close to zero on average so we expect reactive variations to be more prominent. 

\begin{figure*}[t]
    \centering
    \includegraphics[width=1.0\linewidth]{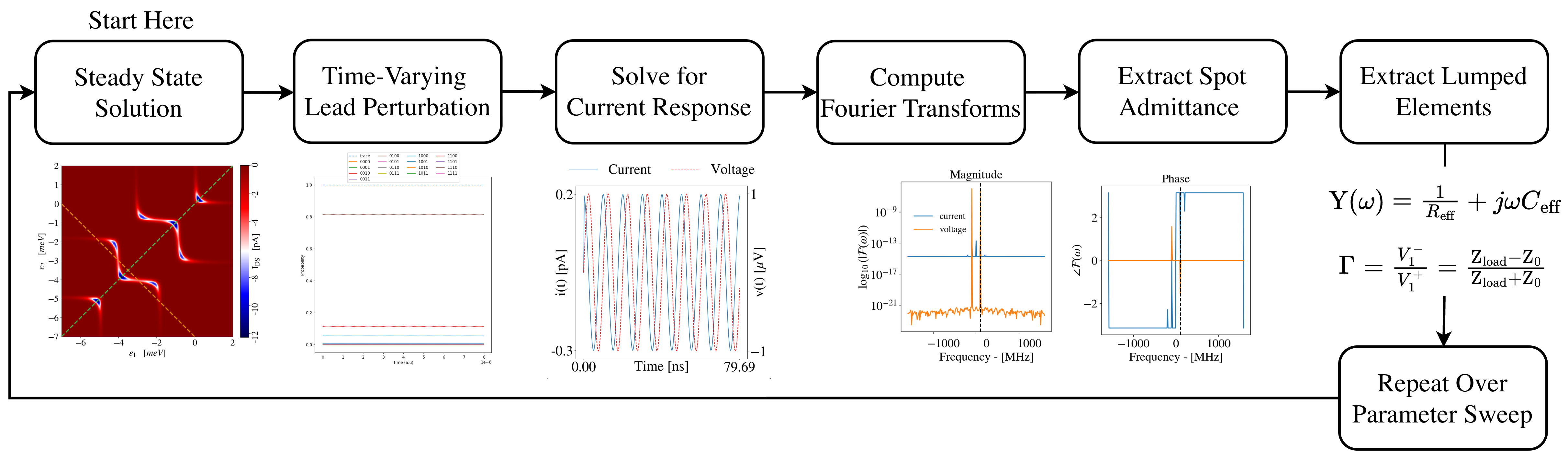}
    \caption{Overview of the proposed reflectometry simulation flow. Beginning with the steady state solution, we then apply a low amplitude, low frequency perturbation to a desired lead. Solving for the time-dependent current response, we can then find the admittance through Fourier transforms of the current and voltage response. This allows us to extract the admittance at the perturbing frequency and finally extract lumped elements such as conductance and effective capacitance. This is then repeated for a sweep over any of the device parameters such as onsite potential.}
    \label{fig:QuantumMeasurement_ReflectoFullFlow}
\end{figure*}

We now present a simulation framework for reflectometry using the semiclassical rate equation. We use the steady state results as a starting point, on top of which we apply a time-varying voltage perturbation on one of the leads. Through monitoring of the time-varying current response we can then define the admittance $Y$ at a frequency $\omega$:
\begin{equation}
    Y(\omega) = \frac{\mathcal{F}(i(t))}{\mathcal{F}(v(t))} = \frac{I(\omega)}{V(\omega)} = G + jB
\end{equation}
where $G$ is conductance, $B$ is susceptance, and the Fourier transform is defined as $\mathcal{F}(x(t)) = X(\omega) = \int_{t=-\infty}^{\infty}x(t)e^{j\omega t}dt$. In the purely real or purely imaginary cases, conductance is related to resistance via $G = R^{-1}$ and susceptance to capacitance via $B = \omega{C}$. However, the admittance is typically complex and so we use conductances and effective capacitances as in \cite{chorley_measuring_2012, frake_radio-frequency_2015}:
\begin{equation}
    Y(\omega) = G + jB = G + j\omega C_{\text{eff}}
\end{equation}

This simulation framework is limited by the perturbation frequency and the amplitude. To ensure the rate equation approach remains valid the frequency must be much lower than any of the energy gaps in the system and the amplitude must not induce unintended tunnelling events between quantum dots. Future work will explore fully coherent dynamics using the Lindblad equation in the frequency domain for fast solving. The full simulation flow in this work for modelling reflectometry using the semiclassical framework is shown in Fig.\,\ref{fig:QuantumMeasurement_ReflectoFullFlow}.

Beginning with the steady state solution, we apply a time-varying perturbation to the lead chemical potential. The time-dependent rate equation is then solved over a number of periods of the perturbation frequency with a high resolution numerical BDF solver in SciPy\,\cite{virtanen_scipy_2020}. 

The experimental probe frequencies in previous works\,\cite{sokolov_common-mode_2025, power_fully-tunable_2025} are $246\,\text{MHz}$ and $75.5\,\text{MHz}$ respectively. So by way of example, we consider a probe frequency of $100\,\text{MHz}$. The RF carrier power incident on the PCB holding the devices is estimated between -70 and -80 dBm in \cite{berritta_real-time_2024}, using a similar experimental setup as in our own works. Therefore we estimate the power reaching the device to be even lower, giving an approximate amplitude of $1\,\mu\text{V}$ at the device. This is in line with an amplitude of $10\mu\text{V}$ in \cite{chorley_measuring_2012}. It is important to ensure the amplitude is not too large so as to invalidate the small perturbation assumption. One method to ensure this is to inspect the Fourier transform of the current and confirm the magnitude of any overtones is much smaller than the probe frequency response. See the supplementary materials for further details.

\section{Reflectometry Results}

Unless otherwise stated we will assume that we are applying the reflectometry perturbation to the source of the quantum dot array. Reflectometry for these idealised systems shows us only the differential admittance. The current does not vary in these simulations within charge stable regions and so there the admittance goes to zero. We also note that despite only extracting admittance due to tunnel rate perturbations in this current work, there is close matching with experimental features.

\subsection{DISET and DISEB}\label{sec:rf_results_disetdiseb}

\begin{figure}[hb]
    \centering
    \subfloat[]{\includegraphics[width=0.48\linewidth]{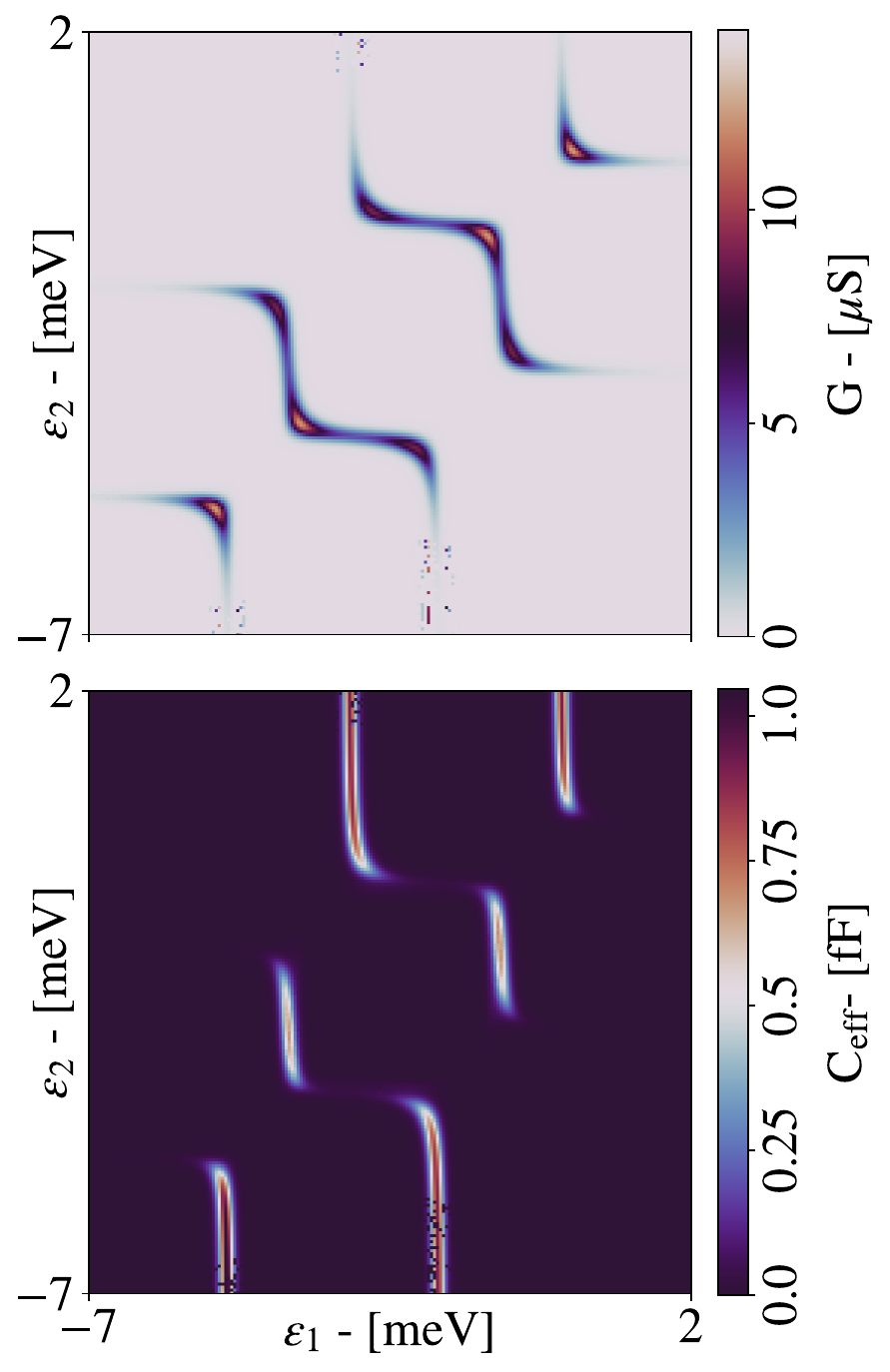}}
    \subfloat[]{\includegraphics[width=0.48\linewidth]{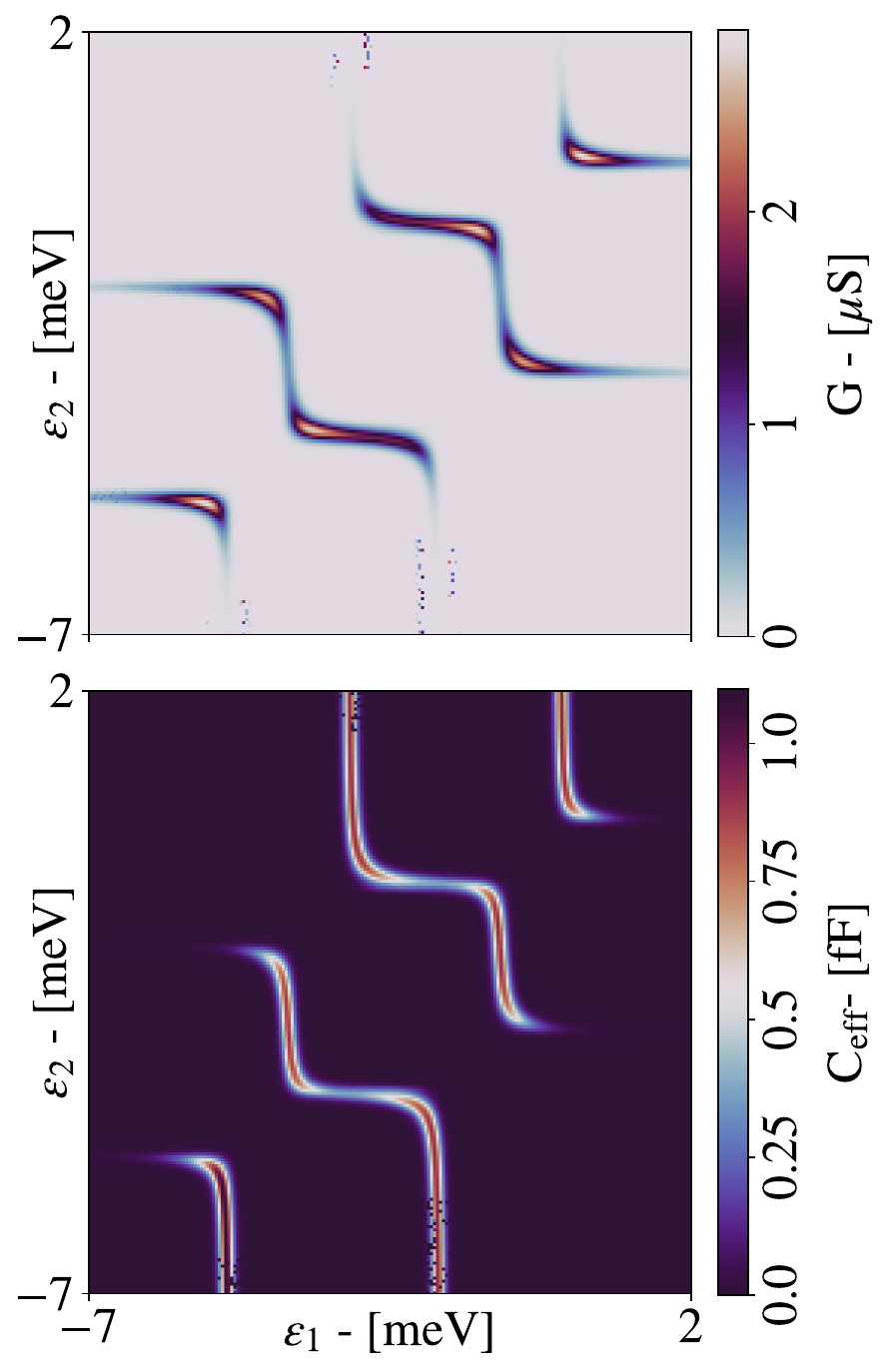}}

    \caption{Conductance and effective capcitance responses for DISET and DISEB. Common Parameters: $U_1 = U_2 = 3.1\,\text{meV}, U_{12} = 0.9\,\text{meV}, t= 200\,\mu\text{eV}, \tau = 10\,\mu\text{eV}, \text{B} = 10\,\text{mT}, \alpha_{1}^{b} = 0.1, \text{T} = 500\,\text{mK}$. (a) DISET Parameters: $\gamma_L =\gamma_R$. (b) DISEB Parameters: $\gamma_R = 0.1\gamma_R$}
    \label{fig:diset_diseb_both}
\end{figure}

We find the DISET conductance and effective capacitance responses in Fig.\,\ref{fig:diset_diseb_both}(a) for onsite potential sweeps. These results show similar charge boundaries as in the steady state current response in the supplementary materials.%
The conductance, in particular in Fig.\,\ref{fig:diset_diseb_both}(a), is a good match with the charge boundaries found in the steady state current case. This is consistent with increases in conductance corresponding to increases in current. However, we now also extract $\text{C}_{\text{eff}}$ for the DISET in Fig.\,\ref{fig:diset_diseb_both}(a). The  asymmetry of sensing via the source reservoir is visible in the capacitance response but not in the conductance. Therefore, both must be viewed together to fully understand the response of the device.

The DISEB simulation results are then shown in Fig.\,\ref{fig:diset_diseb_both}(b). In constrast with the DISET, the conductance response is now no longer symmetric with respect to the source and drain couplings. The conductance is much lower as expected when sensing in box mode. The magnitude of the capacitance response is not increased for the DISEB but does appear enhanced along the transition lines, which is in line with box mode sensing. Note there is some numerical noise visible in these simulation results, which is likely due to the rate equation being inapplicable at these points. Far from interactions we would expect some of the eigenstates to become degenerate and thus invalidate the use of the rate equation at low temperatures. We note in the next section for experimental fitting that this numerical noise is no longer visible at higher temperatures.

\section{Experimental Fitting}
We now consider an example of experimental fitting. We must include the full LC cavity circuit to compare with experiment. As noted, we currently neglect the non-zero classical current flowing through these devices which may have an effect on the final results we present here and the variation in the simulated magnitude and phase. This is a simplified version of the matching circuit used in \cite{power_fully-tunable_2025} and requires further optimisation.

\begin{figure}[htb]
    \centering
    \includegraphics[width=1.0\linewidth]{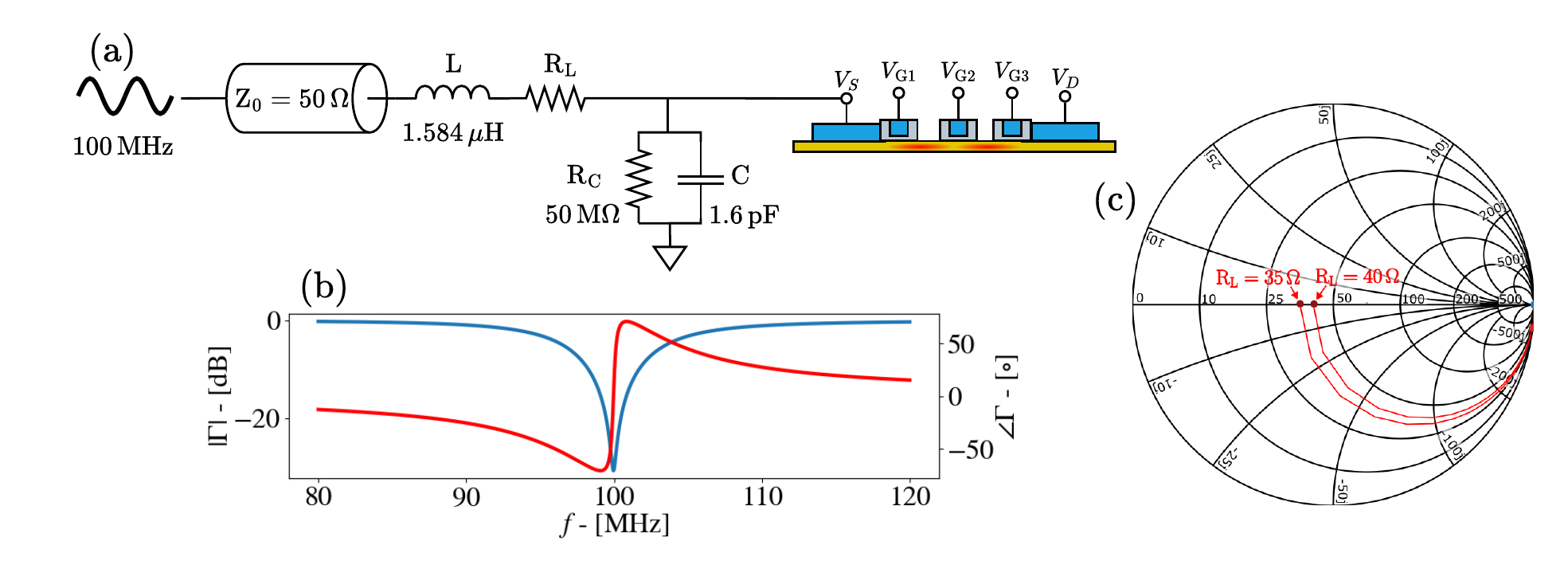}
    \caption{(a) Example matching circuit similar to that used in \cite{power_fully-tunable_2025} connected to the source of a double quantum dot device. Here the circuit is optimised for a 100MHz reflectometry signal. (b) The magnitude and phase response of the LC cavity circuit, showing resonance at 100MHz. (c) Smith chart showing how matching changes with variations in the resistance $\text{R}_{\text{L}}$.}
    \label{fig:QuantumMeasurement_Reflecto_MatchingSetup}
\end{figure}

\subsection{DISET}

\begin{figure}[htb]
    \centering
    \includegraphics[width=1.0\linewidth]{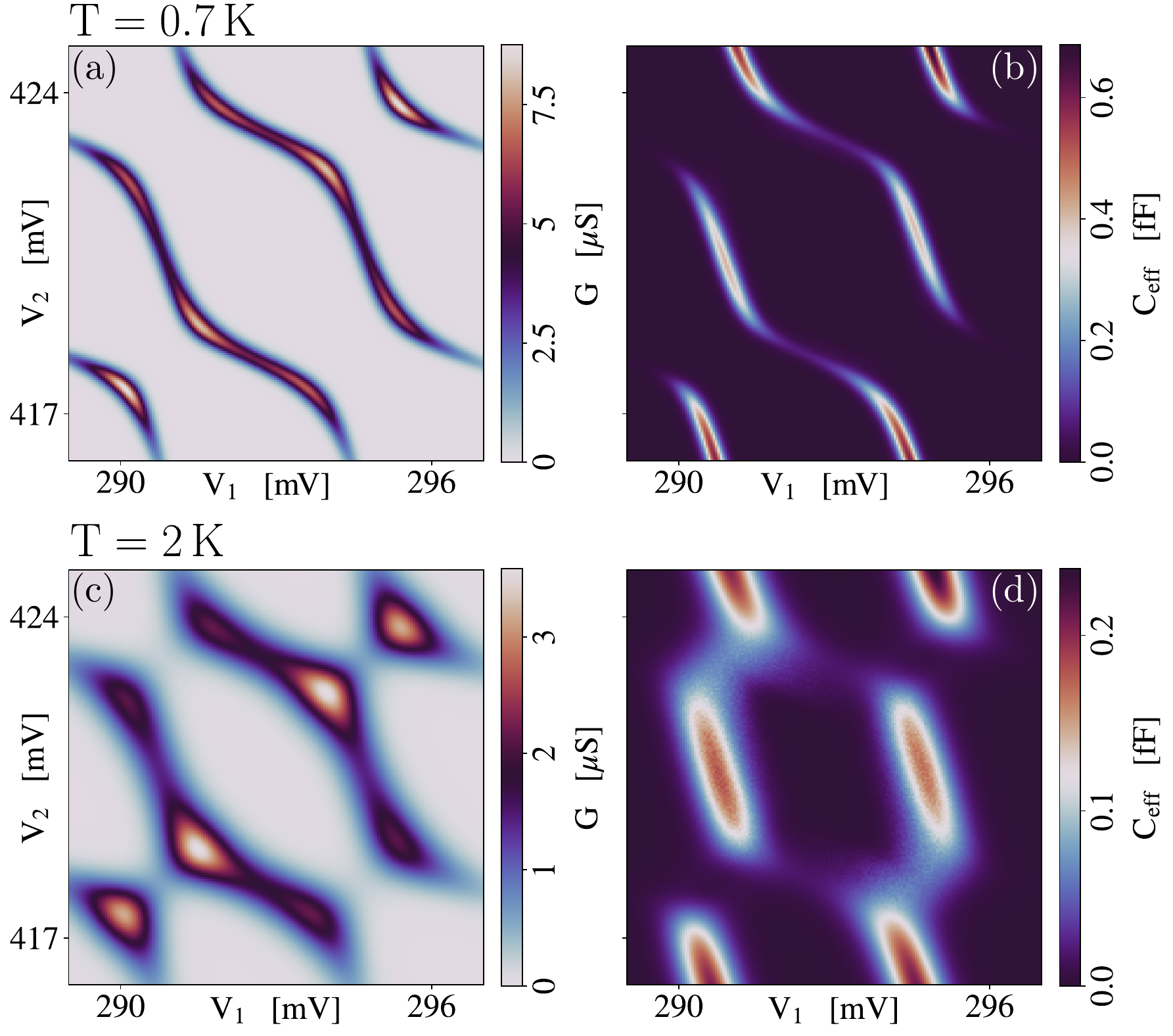}
    \caption{DISET simulation results across temperature. (a),(b) show the conductance and effective capacitance respectively at $T=0.7\,\text{K}$. This is the same temperature that the experimental data in Fig.\,\ref{fig:experimental_measured} was collected at. (c),(d) show the conductance and effective capacitance respectively at $T=2\,\text{K}$. The width of the charge transitions appears to provide a better qualitative match with the experimental data suggesting unintentional broadening in experiment.}
    \label{fig:diset_temp_comparison}
\end{figure}

\begin{figure}[hbt]
    \centering
    \includegraphics[width=1.0\linewidth]{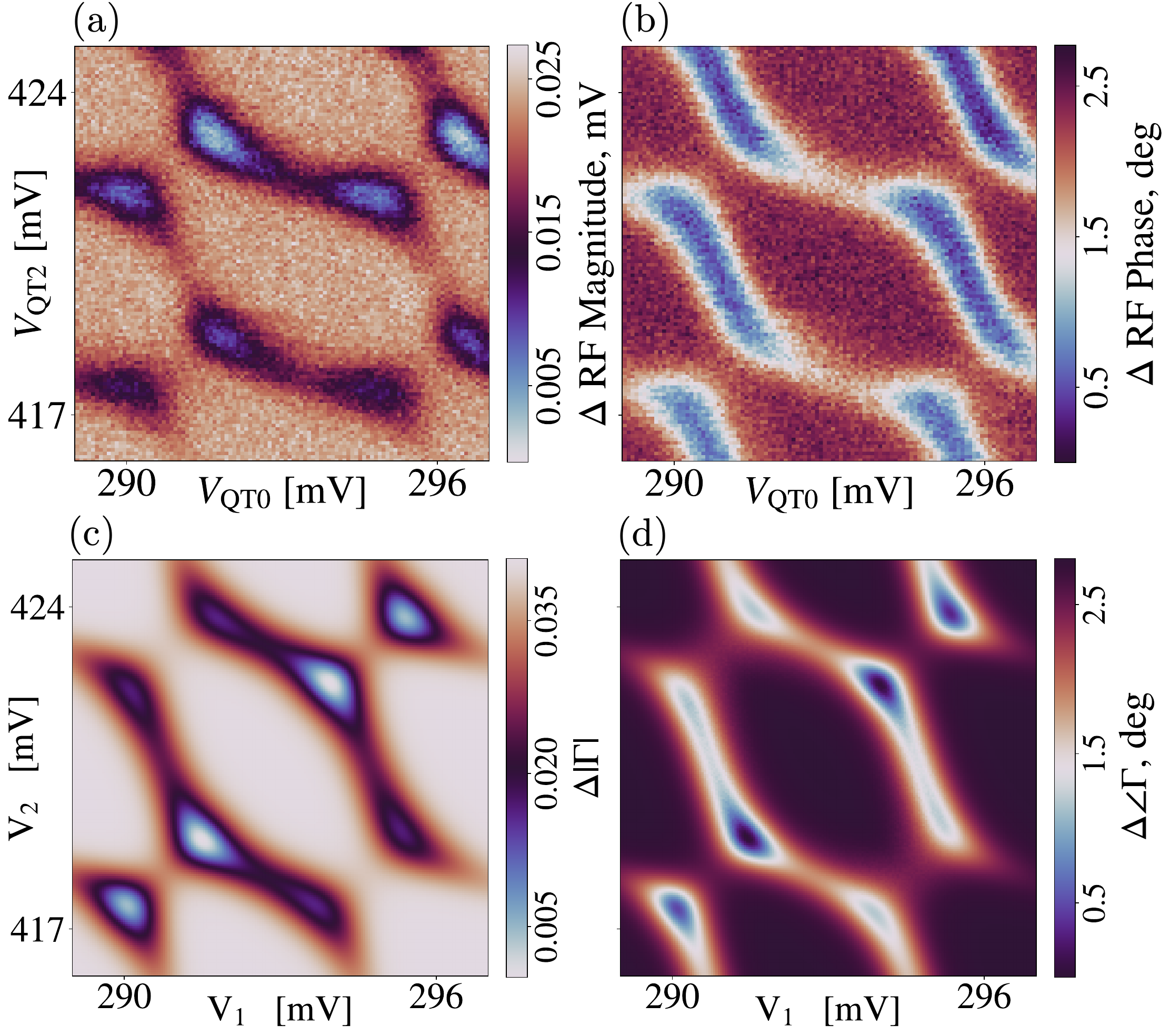}
    \caption{(a),(b) show experimentally measured data for a DISET in GlobalFoundries' 22FDX process\,\cite{power_fully-tunable_2025}. (a) Magnitude of the measured reflection. (b) Phase of the measured reflection. (c),(d) show the simulated reflectometry magnitude and phase response respectively for the DISET for $\text{R}_{\text{L}} = 41\Omega$.}
    \label{fig:experimental_measured}
\end{figure}

We inspect a small section of experimental data\,\cite{power_fully-tunable_2025} for the DISET and compare with simulation. The experimental data in Fig.\,\ref{fig:experimental_measured} was collected at $700\,\text{mK}$ so we first simulate the DISET at the same temperature and record its conductance, G, and effective capacitance, $\text{C}_{\text{eff}}$, in Fig.\,\ref{fig:diset_temp_comparison}\,(a) and (b) respectively.

The extracted energies from experiment are\,\cite{power_fully-tunable_2025}: $(U_1,U_2) = (3, 3.5)~\text{meV}, U_{12} = 0.9~\text{meV}, t = 0.3\,\text{meV}$. The rest of the parameters are taken as: $\tau = 10\,\mu\text{eV}, \gamma_L = \gamma_R, \alpha_{1}^{b} = 0.1, B=10\,\text{mT}.$ The value of the hybridisation is taken from \cite{petropoulos_nanoscale_2024} which also uses the 22FDX process.

 The experimental charge transitions appear significantly more broadened in Fig.\,\ref{fig:experimental_measured} than in the simulation data at $700\,\text{mK}$, so we also show the simulated DISET conductance and capacitance at $2\,\text{K}$ in Fig.\,\ref{fig:diset_temp_comparison}(c) and (d). This broadening is possibly due to the electron temperature being higher than could be experimentally recorded or as a result of the incident power of the RF signal. The rate equation modelling used here cannot faithfully capture broadening due to the RF signal power. Therefore, as an initial model, we increase the temperature of the simulation until there is close matching with experiment. This therefore suggests the electron temperature is in fact higher than could be recorded.

Qualitatively, the simulated data for $\text{T} = 2\,\text{K}$ in Fig.\,\ref{fig:diset_temp_comparison}\,(c) and (d) provide a good match for the charge stability boundaries seen in the experimental data. Most importantly, we capture the asymmetry of sensing via the source reservoir. In experiment we see large phase variations strongly coupled to variations in $V_{\text{QT0}}$, whilst in simulation in Fig.\,\ref{fig:diset_temp_comparison}(d) we see large variations in the effective capacitance coupled to $\text{V}_1$. Note that the RF magnitude appears somewhat asymmetric in experiment as we increase $V_{\text{QT2}}$ as this gate acts as both the plunger and tunnel barrier separating the quantum dot region from the reservoir. Therefore, as $V_{\text{QT2}}$ increases, the hybridisation also increases. We do not model this effect in simulation, and keep the hybridisation constant. 

To more properly compare the simulation outputs with experiment, we embed the simulated admittance in the matching network shown in Fig.\,\ref{fig:QuantumMeasurement_Reflecto_MatchingSetup}. We then plot the magnitude and phase of the reflection coefficient, $\Gamma$, for a series resistance of $\text{R}_{\text{L}} = 41\Omega$ in Fig.\,\ref{fig:experimental_measured}(c),(d). These simulations appear to closely match the experimental data in Fig.\,\ref{fig:experimental_measured}, capturing the shapes of the charge transition boundaries in experiment. The variation in the simulated magnitude is on the order of $0.04$, not dissimilar to $0.02$ in experiment. The phase variation in simulation is approximately $1^{\circ}-2^{\circ}$, depending on the value of $\text{R}_{\text{L}}$. In experiment the phase variation is approximately $2^{\circ}$, again matching quite closely with simulation.

\subsection{DISEB}

\begin{figure}[htb]
    \centering
    \includegraphics[width=1.0\linewidth]{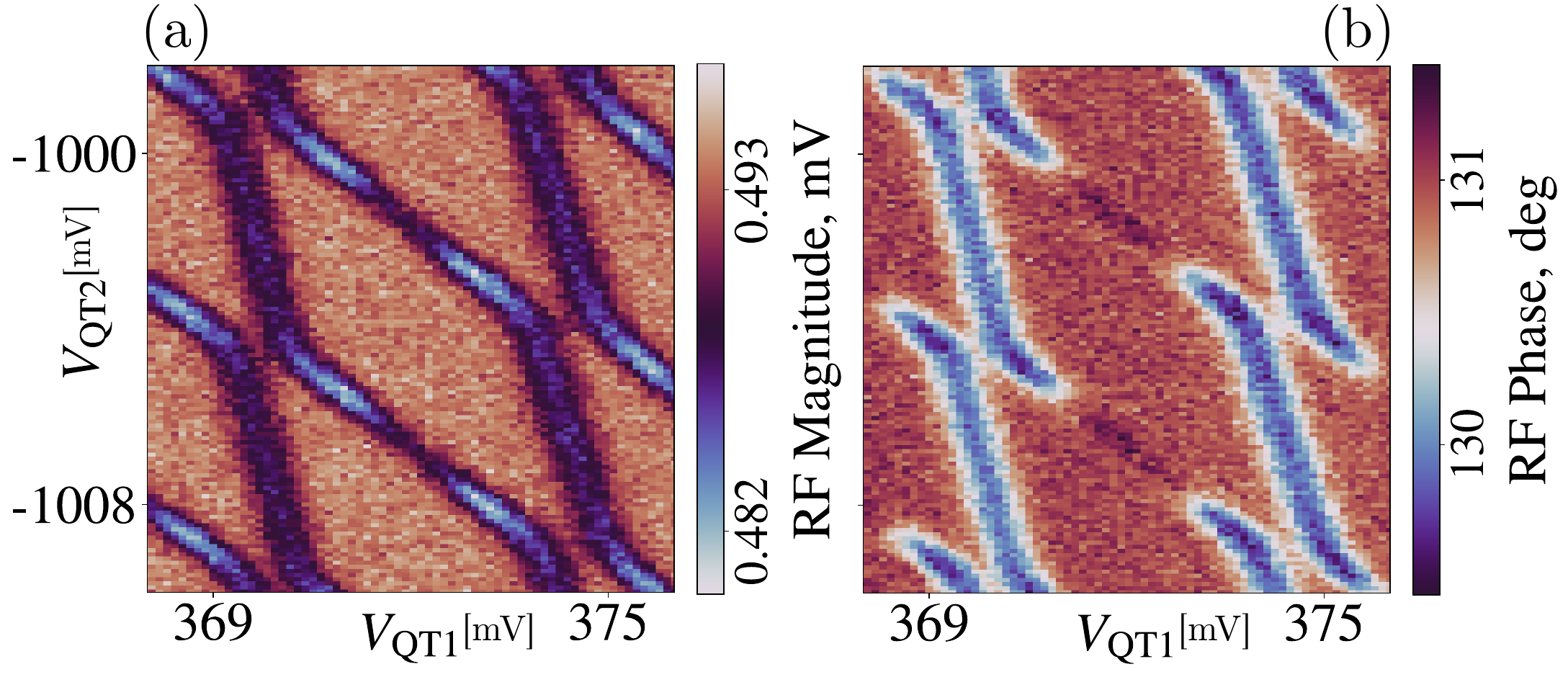}
    \caption{Experimentally measured data for a DISEB in GlobalFoundries' 22FDX process. (a) Magnitude of the measured reflection. (b) Phase of the measured reflection.}
    \label{fig:diseb_experimental}
\end{figure}

\begin{figure}[htb]
    \centering
    \includegraphics[width=1.0\linewidth]{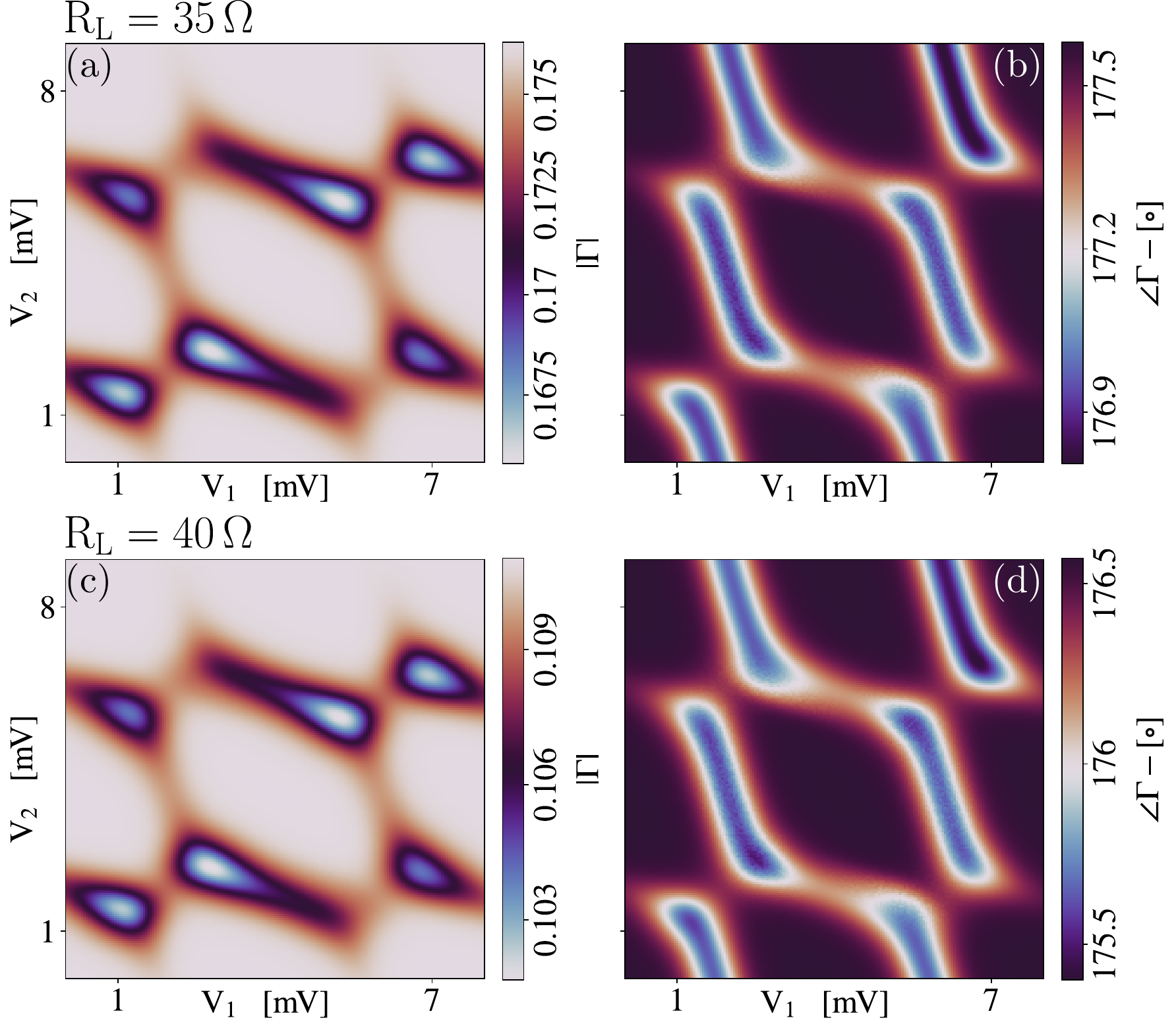}
    \caption{DISEB simulated reflection coefficient results at T = 2\,K. (a),(b) show the magnitude and phase response respectively for $\text{R}_{\text{L}} = 35\,\Omega$. (c),(d) show the magnitude and phase response respectively for $\text{R}_{\text{L}} = 40\,\Omega$.}
    \label{fig:diseb_simulation_RL}
\end{figure}

We now look at experimental data for the DISEB in Fig.\,\ref{fig:diseb_experimental}, similar to that shown in literature\,\cite{power_fully-tunable_2025}. We repeat the same exercise that we did for the DISET and simulate the conductance and effective capacitance variation. However, do note that we are using the same extracted values as we did for the DISET for the DISEB. This was only done in the interest of time and future work aims to repeat these simulations with updated extractions for the DISEB itself. Nonetheless, we still capture many of the important features of the boundaries between charge stable regions in the DISEB in Fig.\,\ref{fig:diseb_simulation_RL}.

We take $\gamma_{\text{D}} = 0.1\gamma_{\text{S}}$ in simulation to capture box-mode sensing and note that this assumption closely aligns simulation results with experimental data. This is a parameter sweep that will be investigated in future work to analyse if we can estimate the level of coupling from experimental data through simulation. A similar value was used in \cite{chorley_measuring_2012} where $\gamma_{\text{R}} = 0.15\gamma_{\text{L}}$.

Two examples of $\text{R}_{L} = 35\Omega \text{ and } 40\Omega$ are shown in the simulation results in Fig.\,\ref{fig:diseb_simulation_RL}. There is not as dramatic variation in the charge stable boundaries as there was in the case of the DISET. However, we do observe some shifting in the range of the magnitude and phase response. In experiment we observe approximately a $1^{\circ}$ change at the charge boundaries, which matches closely with the $1^{\circ}$ change found in simulation for $\text{R}_{L} = 40\,\Omega$. Further work is needed to match the absolute value of the magnitude variations between simulation and experiment, but the relative change is not too dissimilar between the two. Also note there is noise within the charge stable regions in the experimental data that is not present in simulation. This reduces the variation at charge boundaries in experiment versus simulation. Essentially, the boundaries are much clearer in simulation as would expect since we ignore any classical or leakage current through the device. The notes for effects not captured in the DISET are also applicable here in the case of the DISEB. 

The close matching between experimental data for both the DISET and DISEB with the rate equation simulations further confirms this is an appropriate framework for reflectometry modelling in these devices at these high temperatures.

\section{Conclusions}

This work presented a number of results in the area of reflectometry modelling. We employed a rate equation framework to capture the low frequency response of quantum dot arrays. Within a linear response, we could find the admittance, associated conductance, and effective capacitance of the device. This resulted in a number of simulations that closely match experimental data, thus validating the framework.

There are a number of effects not currently captured in simulation but are noted for future work. We aim to increase the number of electrons to avoid edge effects in generating charge stability boundaries. To better compare with experiment we should operate the simulation away from the boundaries of no electrons and the maximum number of electrons allowed. The absolute value of the magnitude and phase do not match experiment exactly. However, this is highly dependent on the matching network used. Therefore, the relative change is of more interest when simulating but this could be optimised for in future simulations. In experiment, for non-zero $\text{V}_{\text{DS}}$, we have observed non-zero current flowing in charge stable regions. Therefore, a more accurate experimental matching may include a classical current contribution in parallel with the quantum transport current. This would add an approximately constant offset across the charge stability plot. We also aim to explore a fully coherent approach using the Lindblad equation in future work. 

\renewcommand{\selectlanguage}[1]{} %
\bibliography{references}

@article{virtanen_scipy_2020,
	title = {{SciPy} 1.0: fundamental algorithms for scientific computing in {Python}},
	volume = {17},
	copyright = {2020 The Author(s)},
	issn = {1548-7105},
	shorttitle = {{SciPy} 1.0},
	url = {https://www.nature.com/articles/s41592-019-0686-2},
	doi = {10.1038/s41592-019-0686-2},
	language = {en},
	number = {3},
	journal = {Nature Methods},
	publisher = {Nature Publishing Group},
	author = {Virtanen, Pauli and Gommers, Ralf and Oliphant, Travis E. and Haberland, Matt and Reddy, Tyler and Cournapeau, David and Burovski, Evgeni and Peterson, Pearu and Weckesser, Warren and Bright, Jonathan and van der Walt, Stéfan J. and Brett, Matthew and Wilson, Joshua and Millman, K. Jarrod and Mayorov, Nikolay and Nelson, Andrew R. J. and Jones, Eric and Kern, Robert and Larson, Eric and Carey, C. J. and Polat, Ilhan and Feng, Yu and Moore, Eric W. and VanderPlas, Jake and Laxalde, Denis and Perktold, Josef and Cimrman, Robert and Henriksen, Ian and Quintero, E. A. and Harris, Charles R. and Archibald, Anne M. and Ribeiro, Antonio H. and Pedregosa, Fabian and van Mulbregt, Paul},
	month = mar,
	year = {2020},
	pages = {261--272},
}

@article{peri_unified_2024,
	title = {Unified linear response theory of quantum electronic circuits},
	volume = {10},
	copyright = {2024 The Author(s)},
	issn = {2056-6387},
	url = {https://www.nature.com/articles/s41534-024-00907-9},
	doi = {10.1038/s41534-024-00907-9},
	language = {en},
	number = {1},
	journal = {npj Quantum Information},
	publisher = {Nature Publishing Group},
	author = {Peri, L. and Benito, M. and Ford, C. J. B. and Gonzalez-Zalba, M. F.},
	month = nov,
	year = {2024},
	pages = {1--14},
}

@article{petropoulos_nanoscale_2024,
	title = {Nanoscale single-electron box with a floating lead for quantum sensing: {Modeling} and device characterization},
	volume = {124},
	issn = {0003-6951, 1077-3118},
	shorttitle = {Nanoscale single-electron box with a floating lead for quantum sensing},
	url = {https://pubs.aip.org/apl/article/124/17/173503/3284587/Nanoscale-single-electron-box-with-a-floating-lead},
	doi = {10.1063/5.0203421},
	language = {en},
	number = {17},
	journal = {Applied Physics Letters},
	author = {Petropoulos, N. and Wu, X. and Sokolov, A. and Giounanlis, P. and Bashir, I. and Mitchell, A. K. and Asker, M. and Leipold, D. and Staszewski, R. B. and Blokhina, E.},
	month = apr,
	year = {2024},
	pages = {173503},
}

@misc{noauthor_qtcad_nodate,
	title = {{QTCAD}, https://nanoacademic.com/solutions/qtcad/},
	url = {https://nanoacademic.com/solutions/qtcad/},
}

@article{mukherjee_machine_2025,
	title = {Machine learning unveils multiple {Pauli} blockades in the transport spectroscopy of bilayer graphene double-quantum dots},
	volume = {2},
	issn = {2835-0103},
	url = {https://doi.org/10.1063/5.0261094},
	doi = {10.1063/5.0261094},
	number = {2},
	journal = {APL Quantum},
	author = {Mukherjee, Ankan and Das, Anuranan and Khan, Adil Anwar and Muralidharan, Bhaskaran},
	month = may,
	year = {2025},
	pages = {026125},
}

@article{peri_compact_2025,
	title = {Compact quantum dot models for analog microwave co-simulation},
	volume = {11},
	copyright = {2025 The Author(s)},
	issn = {2056-6387},
	url = {https://www.nature.com/articles/s41534-025-01140-8},
	doi = {10.1038/s41534-025-01140-8},
	language = {en},
	number = {1},
	journal = {npj Quantum Information},
	publisher = {Nature Publishing Group},
	author = {Peri, Lorenzo and Gomez-Saiz, Alberto and Ford, Christopher J. B. and Gonzalez-Zalba, M. Fernando},
	month = dec,
	year = {2025},
	pages = {194},
}

@article{cochrane_intrinsic_2024,
	title = {Intrinsic noise of the single-electron box},
	volume = {21},
	issn = {2331-7019},
	url = {https://link.aps.org/doi/10.1103/PhysRevApplied.21.064066},
	doi = {10.1103/PhysRevApplied.21.064066},
	language = {en},
	number = {6},
	journal = {Physical Review Applied},
	author = {Cochrane, Laurence and Seshia, Ashwin A. and Gonzalez-Zalba, M. Fernando},
	month = jun,
	year = {2024},
	pages = {064066},
}

@article{swift_superinductor_2025,
	title = {A superinductor in a deep sub-micron integrated circuit},
	url = {http://arxiv.org/abs/2507.13202},
	journal = {arXiv:2507.13202},
	publisher = {arXiv},
	author = {Swift, T. H. and Olivieri, F. and Aizpurua-Iraola, G. and Kirkman, J. and Noah, G. M. and Kruijf, M. de and Horstig, F. E. von and Gomez-Saiz, A. and Morton, J. J. L. and Gonzalez-Zalba, M. F.},
	month = nov,
	year = {2025},
}

@article{sokolov_common-mode_2025,
	title = {Common-mode control and confinement inversion of electrostatically defined quantum dots in a commercial {CMOS} process},
	copyright = {2025 The Author(s)},
	issn = {2045-2322},
	url = {https://www.nature.com/articles/s41598-025-28278-x},
	doi = {10.1038/s41598-025-28278-x},
	language = {en},
	journal = {Scientific Reports},
	publisher = {Nature Publishing Group},
	author = {Sokolov, Andrii and Wu, Xutong and Power, Conor and Asker, Mike and Giounanlis, Panagiotis and Kriekouki, Ioanna and Hanos-Puskai, Pèter and McGeough, Conor and Bashir, Imran and Redmond, David and Leipold, Dirk and Staszewski, Robert Bogdan and Blokhina, Elena},
	month = nov,
	year = {2025},
}

@article{ahmed_primary_2018,
	title = {Primary thermometry of a single reservoir using cyclic electron tunneling to a quantum dot},
	volume = {1},
	copyright = {2018 The Author(s)},
	issn = {2399-3650},
	url = {https://www.nature.com/articles/s42005-018-0066-8},
	doi = {10.1038/s42005-018-0066-8},
	language = {en},
	number = {1},
	journal = {Communications Physics},
	publisher = {Nature Publishing Group},
	author = {Ahmed, Imtiaz and Chatterjee, Anasua and Barraud, Sylvain and Morton, John J. L. and Haigh, James A. and Gonzalez-Zalba, M. Fernando},
	month = oct,
	year = {2018},
	pages = {1--7},
}

@article{frake_radio-frequency_2015,
	title = {Radio-frequency capacitance spectroscopy of metallic nanoparticles},
	volume = {5},
	issn = {2045-2322},
	url = {https://www.nature.com/articles/srep10858},
	doi = {10.1038/srep10858},
	language = {en},
	number = {1},
	journal = {Scientific Reports},
	author = {Frake, James C. and Kano, Shinya and Ciccarelli, Chiara and Griffiths, Jonathan and Sakamoto, Masanori and Teranishi, Toshiharu and Majima, Yutaka and Smith, Charles G. and Buitelaar, Mark R.},
	month = jun,
	year = {2015},
	pages = {10858},
}

@book{schaller_open_2014,
	address = {Cham},
	series = {Lecture {Notes} in {Physics}},
	title = {Open {Quantum} {Systems} {Far} from {Equilibrium}},
	volume = {881},
	copyright = {https://www.springernature.com/gp/researchers/text-and-data-mining},
	isbn = {978-3-319-03876-6 978-3-319-03877-3},
	url = {https://link.springer.com/10.1007/978-3-319-03877-3},
	doi = {10.1007/978-3-319-03877-3},
	language = {en},
	publisher = {Springer International Publishing},
	author = {Schaller, Gernot},
	year = {2014},
}

@article{kirsanskas_qmeq_2017,
	title = {{QmeQ} 1.0: {An} open-source {Python} package for calculations of transport through quantum dot devices},
	volume = {221},
	issn = {00104655},
	shorttitle = {{QmeQ} 1.0},
	url = {https://linkinghub.elsevier.com/retrieve/pii/S0010465517302515},
	doi = {10.1016/j.cpc.2017.07.024},
	language = {en},
	journal = {Computer Physics Communications},
	author = {Kiršanskas, Gediminas and Pedersen, Jonas Nyvold and Karlström, Olov and Leijnse, Martin and Wacker, Andreas},
	month = dec,
	year = {2017},
	pages = {317--342},
}

@article{chorley_measuring_2012,
	title = {Measuring the complex admittance of a carbon nanotube double quantum dot},
	volume = {108},
	issn = {0031-9007, 1079-7114},
	url = {http://arxiv.org/abs/1109.1827},
	doi = {10.1103/PhysRevLett.108.036802},
	language = {en},
	number = {3},
	journal = {Physical Review Letters},
	author = {Chorley, S. J. and Wabnig, J. and Penfold-Fitch, Z. V. and Petersson, K. D. and Frake, J. and Smith, C. G. and Buitelaar, M. R.},
	month = jan,
	year = {2012},
	pages = {036802},
}

@article{campaioli_tutorial_2023,
	title = {A {Tutorial} on {Quantum} {Master} {Equations}: {Tips} and tricks for quantum optics, quantum computing and beyond},
	shorttitle = {A {Tutorial} on {Quantum} {Master} {Equations}},
	url = {http://arxiv.org/abs/2303.16449},
	journal = {arXiv:2303.16449},
	language = {en},
	publisher = {arXiv},
	author = {Campaioli, Francesco and Cole, Jared H. and Hapuarachchi, Harini},
	month = mar,
	year = {2023},
}

@article{fransson_pauli_2006,
	title = {Pauli spin blockade in weakly coupled quantum dots},
	volume = {73},
	issn = {1098-0121, 1550-235X},
	url = {http://arxiv.org/abs/cond-mat/0509482},
	doi = {10.1103/PhysRevB.73.205333},
	language = {en},
	number = {20},
	journal = {Physical Review B},
	author = {Fransson, J. and Rasander, M.},
	month = may,
	year = {2006},
	pages = {205333},
}

@article{huang_high-sensitivity_2021,
	title = {A {High}-{Sensitivity} {Charge} {Sensor} for {Silicon} {Qubits} above 1 {K}},
	volume = {21},
	issn = {1530-6984, 1530-6992},
	url = {https://pubs.acs.org/doi/10.1021/acs.nanolett.1c01003},
	doi = {10.1021/acs.nanolett.1c01003},
	language = {en},
	number = {14},
	journal = {Nano Letters},
	author = {Huang, Jonathan Yue and Lim, Wee Han and Leon, Ross C. C. and Yang, Chih Hwan and Hudson, Fay E. and Escott, Christopher C. and Saraiva, Andre and Dzurak, Andrew S. and Laucht, Arne},
	month = jul,
	year = {2021},
	pages = {6328--6335},
}

@article{berritta_real-time_2024,
	title = {Real-time two-axis control of a spin qubit},
	volume = {15},
	copyright = {2024 The Author(s)},
	issn = {2041-1723},
	url = {https://www.nature.com/articles/s41467-024-45857-0},
	doi = {10.1038/s41467-024-45857-0},
	language = {en},
	number = {1},
	journal = {Nature Communications},
	publisher = {Nature Publishing Group},
	author = {Berritta, Fabrizio and Rasmussen, Torbjørn and Krzywda, Jan A. and van der Heijden, Joost and Fedele, Federico and Fallahi, Saeed and Gardner, Geoffrey C. and Manfra, Michael J. and van Nieuwenburg, Evert and Danon, Jeroen and Chatterjee, Anasua and Kuemmeth, Ferdinand},
	month = feb,
	year = {2024},
	pages = {1676},
}

@article{power_fully-tunable_2025,
	title = {Fully-{Tunable} {Tunnel}-{Coupled} {Quantum} {Dots} and {Charge} {Sensing} in a {Commercial} 22nm {FD}-{SOI} {Process}},
	issn = {1558-0563},
	url = {https://ieeexplore.ieee.org/document/11107337},
	doi = {10.1109/LED.2025.3595384},
	journal = {IEEE Electron Device Letters},
	author = {Power, C. and Moras, M. and Sokolov, A. and Rohrbacher, C. and Wu, X. and Amitonov, S.V. and Kriekouki, I. and Aprà, A. and Giounanlis, P. and Asker, M. and Harkin, M. and Hanos-Puskai, P. and Bisiaux, P. and Bashir, I. and Redmond, D. and Leipold, D. and Staszewski, R.B. and Barry, B. and Samkharadze, N. and Blokhina, E.},
	year = {2025},
	pages = {1--1},
}

@article{vigneau_probing_2022,
	title = {Probing quantum devices with radio-frequency reflectometry},
	url = {http://arxiv.org/abs/2202.10516},
	journal = {arXiv:2202.10516},
	publisher = {arXiv},
	author = {Vigneau, Florian and Fedele, Federico and Chatterjee, Anasua and Reilly, David and Kuemmeth, Ferdinand and Gonzalez-Zalba, Fernando and Laird, Edward and Ares, Natalia},
	month = feb,
	year = {2022},
}

@book{ihn_semiconductor_2015,
	address = {Oxford New York},
	edition = {Reprinted with corrections},
	title = {Semiconductor nanostructures: quantum states and electronic transport},
	isbn = {978-0-19-953442-5 978-0-19-953443-2},
	shorttitle = {Semiconductor nanostructures},
	language = {en},
	publisher = {Oxford University Press},
	author = {Ihn, Thomas},
	year = {2015},
}

@article{power_charge-based_2024,
	title = {Charge-{Based} {Single}-{Shot} {Readout} {Systems} in {Quantum} {Computing} [{Feature}]},
	volume = {24},
	issn = {1558-0830},
	url = {https://ieeexplore.ieee.org/document/10753314/?arnumber=10753314},
	doi = {10.1109/MCAS.2024.3401618},
	number = {4},
	journal = {IEEE Circuits and Systems Magazine},
	author = {Power, Conor and Staszewski, Robert Bogdan and Blokhina, Elena},
	year = {2024},
	pages = {42--64},
}

@book{bruus_many-body_2016,
	address = {Oxford New York},
	edition = {Corrected version},
	title = {Many-body quantum theory in condensed matter physics: an introduction},
	isbn = {978-0-19-856633-5},
	shorttitle = {Many-body quantum theory in condensed matter physics},
	language = {eng},
	publisher = {Oxford University Press},
	author = {Bruus, Henrik and Flensberg, Karsten},
	year = {2016},
}

\appendix

\newpage
\section{Steady-State Results}\label{sec:stationaryresults}

Some of the results in this section are not necessarily novel in their own right but are important to demonstrate the working of this simulation framework for known results. The results are consistent with those shown in literature\cite{kirsanskas_qmeq_2017,mukherjee_machine_2025}. In \cite{mukherjee_machine_2025} they consider a number of extra effects that we plan to include in future work; interdot spin-flip tunnelling, spin-orbit coupling, valley Zeeman effect, and phonon-assisted tunnelling. We do not consider time-domain simulations in this section, solving only Eq.\,\ref{eqn:steady_state_sol}, and take the probe amplitude applied to the leads, $V_{ac}^{b} = 0$. We will make this non-zero in the following section when we discuss reflectometry and solve the time-dependent rate equation. The parameters used in the simulations are based on Fermi-Hubbard extractions for the 22FDX process\,\cite{power_fully-tunable_2025}. The parameters for each simulation are noted in the caption of each of the plots. The magnetic field is kept non-zero to reduce degeneracies in the system so that the rate-equation approach remains valid.

\subsection{Single-Island Single-Electron Transistor}

\begin{figure}[!htb]
    \centering
    \includegraphics[width=0.9\linewidth]{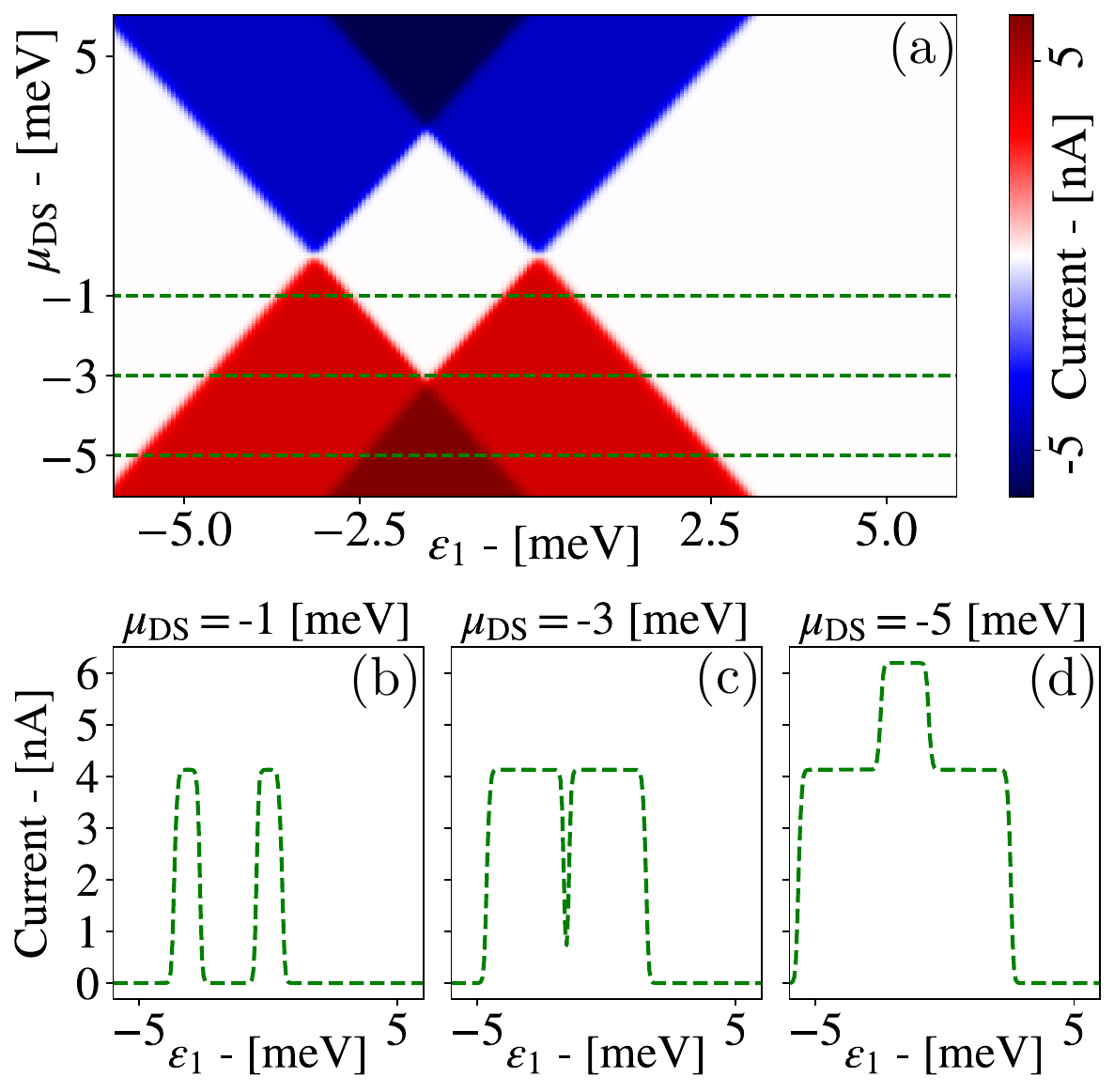}
    \caption{SISET steady state current simulation results. (a) Formation of a Coulomb diamond with a sweep over the onsite potential $\epsilon_1$ and the bias $\mu_{DS}$. (b),(c), and (d) show three linecuts at $\mu_{\text{DS}} = -1, -3, -5\,\text{meV}$ respetively. Parameters: $U_1=3.1\,\text{meV}, \text{B}=1\,\mu\text{T}, \tau = 10\mu\text{eV}, \gamma_L = \gamma_R, \text{T} = 500\,\text{mK}$.}
    \label{fig:siset_coulomb_diamonds_1dot}
\end{figure}

The single-island single-electron transistor (SISET) consists of two electron reservoirs and a single quantum dot in the centre, as shown in Fig.\,\ref{fig:QuantumMeasurement_ReflectoDevices}(a). Here we consider only the limited charge basis set of two electrons with spin, for a total of four charge basis states. We consider a sweep over the onsite potential $\epsilon_1$ and over the drain-source chemical potential difference, $\mu_{\text{DS}} = \mu_{\text{D}} - \mu_{\text{S}}$, in Fig.\,\ref{fig:siset_coulomb_diamonds_1dot}.

We observe the formation of a Coulomb diamond as expected. Slices of constant $\mu_{\text{DS}}$ are also shown in the same figure to illustrate Coulomb peaks and the increase of current through the SISET for a large enough bias window, such as for $\mu_{\text{DS}} = -5\,\text{meV}$. Note the approximate calibration and fitting done for the model input parameters results in a current value in line with what we may expect to measure in experiment, of the order of nanoamperes\,\cite{huang_high-sensitivity_2021}. Now removing one of the reservoirs, we next look at the single-island single-electron box.

\subsection{Single-Island Single-Electron Box}
The single electron box mode we define in the same way as in \cite{power_fully-tunable_2025}, where we disconnect one of the reservoirs, sending the dot/reservoir coupling towards zero for that lead. Therefore, we can no longer probe the current through this device as current does not flow. However, we can still monitor the change in current as electrons hop in and out of the dot region during sequential tunnelling events. Here we show a sweep over the chemical potential of the reservoir that is still connected and then a sweep over the onsite potential of the quantum dot. In each case we keep all other parameters constant. This is similar to what was discussed in \cite{petropoulos_nanoscale_2024} where they consider more electrons in the dot. 
\begin{figure}[!htb]
    \centering
    \includegraphics[width=1.0\linewidth]{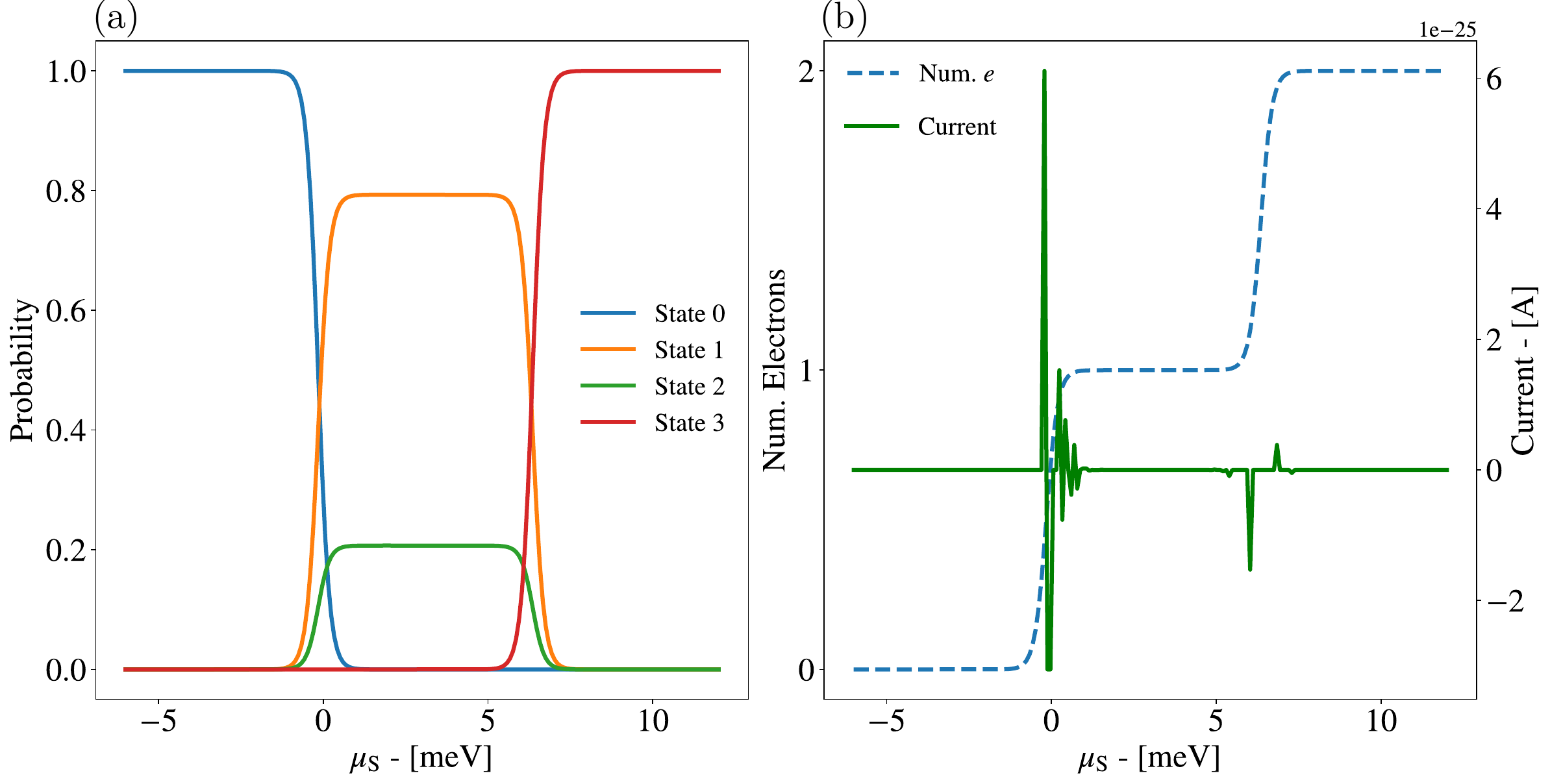}
    \caption{Loading electrons into a single quantum dot through a chemical potential sweep of the tunnel-coupled reservoir. (a) shows the dot populations. (b) shows the expectation of the dot occupation and the current through the source reservoir. Parameters: $\epsilon_1=0, U_1=3.1\,\text{meV}, \text{B}=1\,\mu\text{T}, \tau = 10\mu\text{eV}, \gamma_R = 0, \text{T} = 1\,\text{K}$ }
    \label{fig:siseb_chem_states_current_loading}
\end{figure}

We first show loading electrons through sweeping the chemical potential of the lead in Fig.\,\ref{fig:siseb_chem_states_current_loading}. Notice the sudden current responses at the transitions between charge subspaces as electrons move between the lead and dot. The magnitude of the current is extremely small as electrons only briefly tunnel between the dot and lead regions, with charge transitions in and out of the dot region remaining approximately equal as there is no bias window. We show the current here as it indicates when tunnelling occurs.  Later, in Sec.\,\ref{sec:rf_results_disetdiseb}, we consider $\gamma_{\text{R}} = 0.1\gamma_{\text{L}}$ for reflectometry modelling of box sensors which allows for a larger current to flow during tunnelling events. This is more inline with what we observe in experiment than considering $\gamma_{\text{R}} = 0$, as we do in this current section.
\begin{figure}[!htb]
    \centering
    \includegraphics[width=1.0\linewidth]{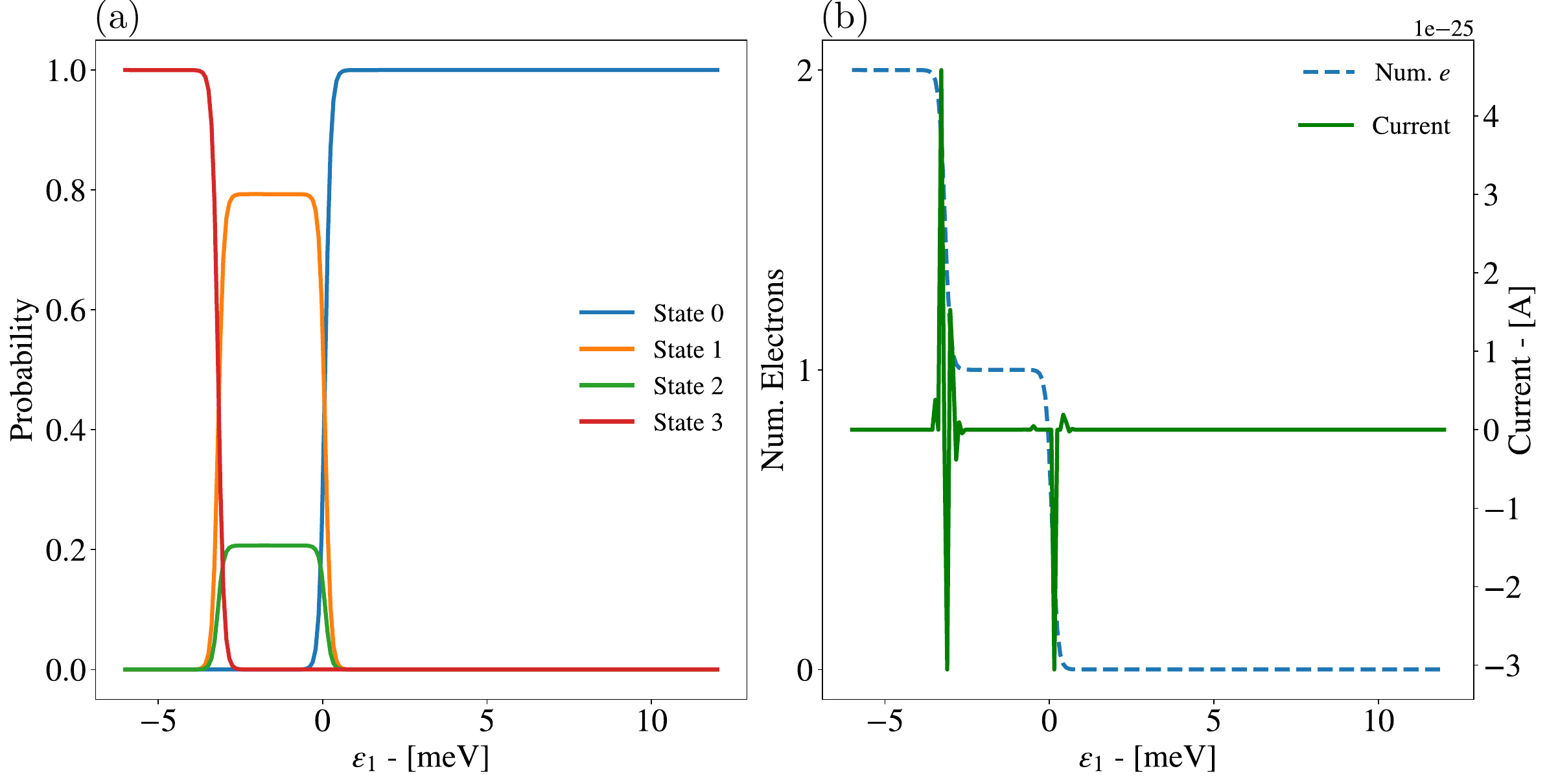}
    \caption{Removing electrons from a single quantum dot tunnel-coupled to an electron reservoir through an onsite sweep. (a) shows the dot charge populations. (b) shows the expectation of the dot occupation and the current through the source reservoir. Parameters are the same as in Fig.\,\ref{fig:siseb_chem_states_current_loading}.}
    \label{fig:siseb_onsite_states_current_loading}
\end{figure}
We then show a similar simulation in Fig.\,\ref{fig:siseb_onsite_states_current_loading}. However, now we remove electrons from the quantum dot with a sweep of the onsite potential, with a constant chemical potential on the reservoir. We also show the populations in  Fig.\,\ref{fig:siseb_onsite_states_current_loading}(a), cycling in the opposite way to the populations in Fig.\,\ref{fig:siseb_chem_states_current_loading}(a) as we would expect, given the sweep over the onsite potential.

\subsection{Double-Island Single-Electron Transistor}
The double-island single-electron transistor (DISET) has two quantum dots in its dot region, connected via a tunable tunnel-coupling $t$. Accounting for up to two electrons on each quantum dot with spin, there are now 16 many-body occupation basis states - see Tab.\,\ref{tab:4-electron-subspaces}.

\begin{figure}[!htb]
    \centering
    \includegraphics[width=0.6\linewidth]{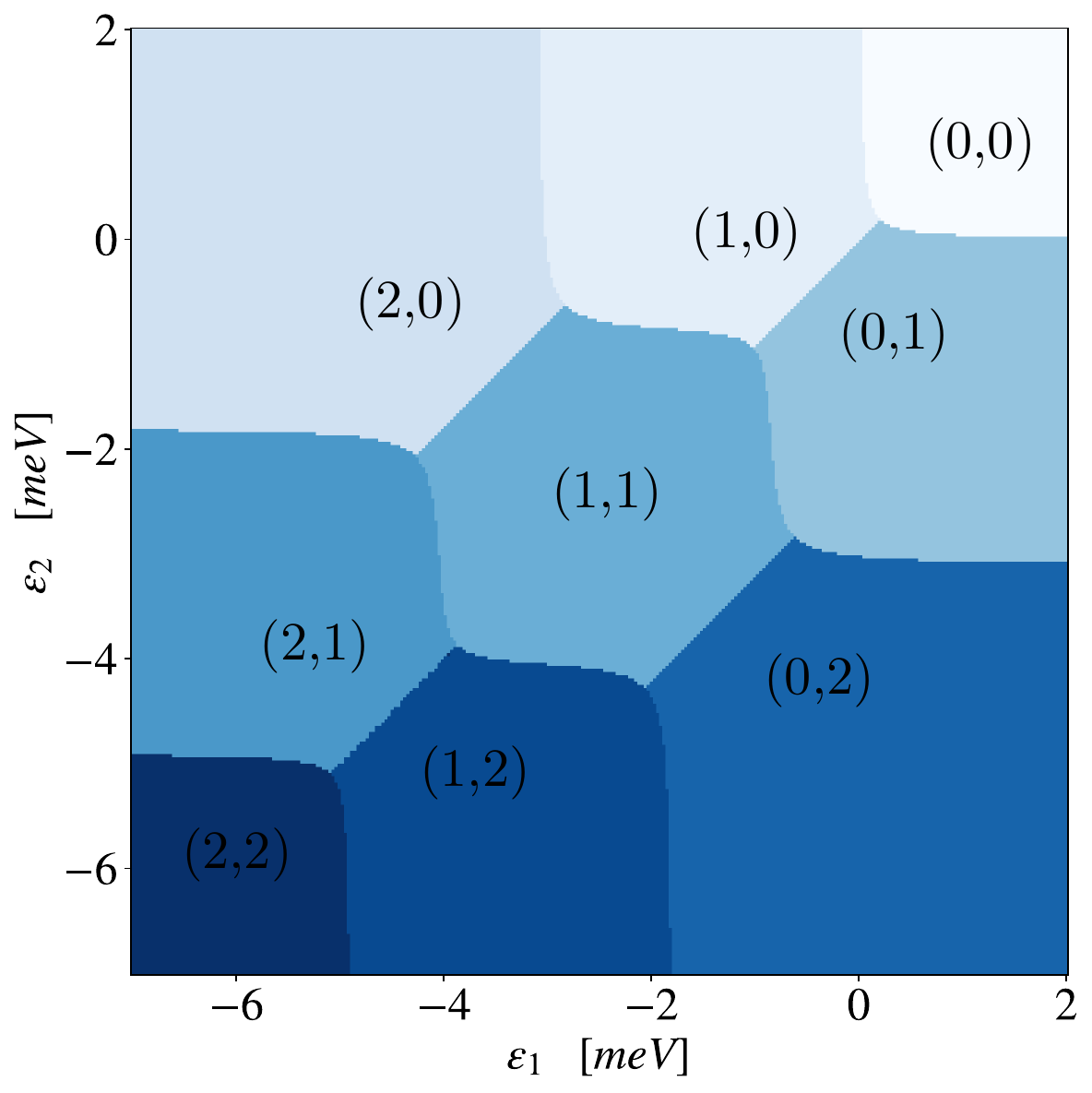}
    \caption{An ideal charge stability diagram derived from the Hamiltonian of the double quantum dot. There is no knowledge of the environment or leads in this plot. Each colour represents a distinct charge configuration, where $(n_1,n_2)$ means there are $n_1$ electrons in the left quantum dot and $n_2$ electrons in the right quantum dot.}
    \label{fig:dd_charge_stability_ideal}
\end{figure}

Sweeping independently over the onsite potentials, $\epsilon_1$ and $\epsilon_2$, we produce the ideal charge stability diagram in Fig.\,\ref{fig:dd_charge_stability_ideal} directly from the ground state of the Hamiltonian $\hat{H}_{\text{S}}$. Each region is charge stable, in the sense that it has a fixed number of electrons in each dot. Increasing the onsite potentials empties the two quantum dots, with the (0,0) occupation in the top right corner of the figure.

\begin{figure}[!htb]
    \centering
    \includegraphics[width=1.0\linewidth]{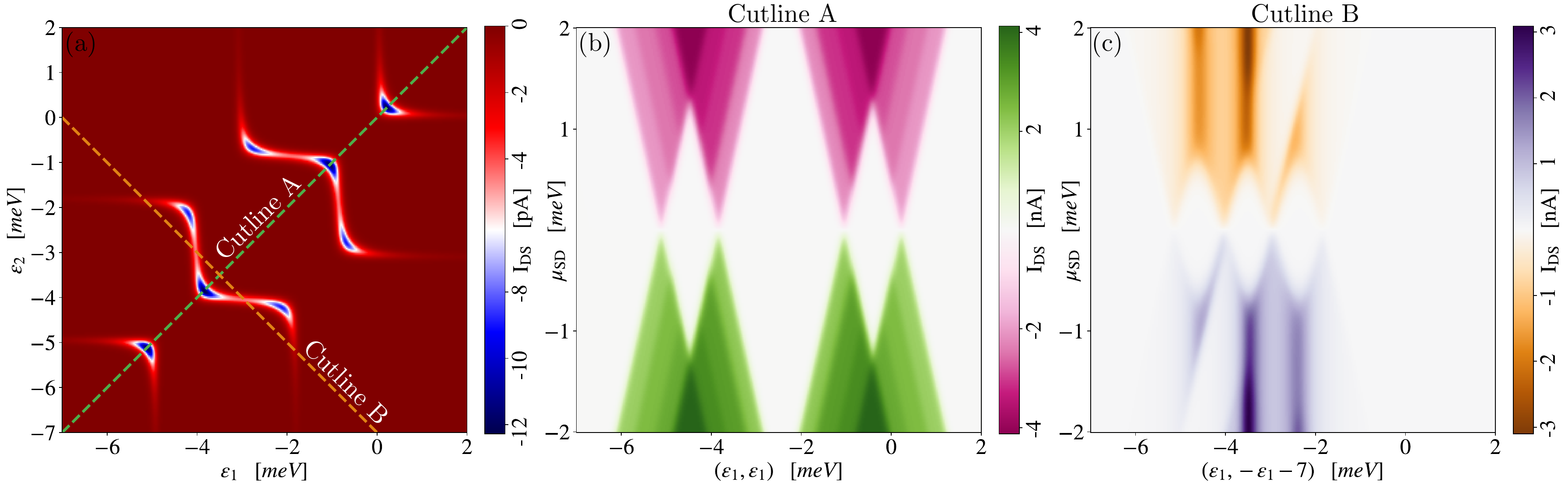}
    \caption{Common Parameters: $U_1 = U_2 = 3.1\,\text{meV}, U_{12} = 0.9\,\text{meV}, t= 200\,\mu\text{eV}, \tau = 10\,\mu\text{eV}, \gamma_L =\gamma_R, \text{T} = 500\,\text{mK}$. (a) DISET current response for onsite potential sweep. The charge stable regions from Fig.\,\ref{fig:dd_charge_stability_ideal} are visible through the current response. Parameters: $\mu_{\text{SD}} = 1\,\mu\text{eV}$ (b) Detuning sweep along cutline A with a bias sweep $\mu_{\text{SD}}$. Note the formation of Coulomb diamonds - there are three visible as we can load up to a maximum of four electrons along this line. (c) Detuning sweep along cutline B with a bias sweep $\mu_{\text{SD}}$. The response is markedly different to cutline A. See Fig.\,\ref{fig:diset_4_charge_stability_PMVds} for further insight.}
    \label{fig:diset_4_charge_stability_vdsSweeps}
\end{figure}

If we now monitor the current through the DISET, we can detect transitions between charge stable regions differing in the number of electrons. Charge stable regions that do not differ by electron count, such as $(2,1) \leftrightarrow (1,2)$, do not result in a net transfer of electrons with the leads and so display no measurable current. However, later in this chapter we will see such transitions are detectable via reflectometry. For $\mu_{\text{DS}} = 1\,\mu\text{eV}$, we simulate the current through a DISET in Fig.\,\ref{fig:diset_4_charge_stability_vdsSweeps}(a).

\begin{figure}[!htb]
    \centering
    \includegraphics[width=1.0\linewidth]{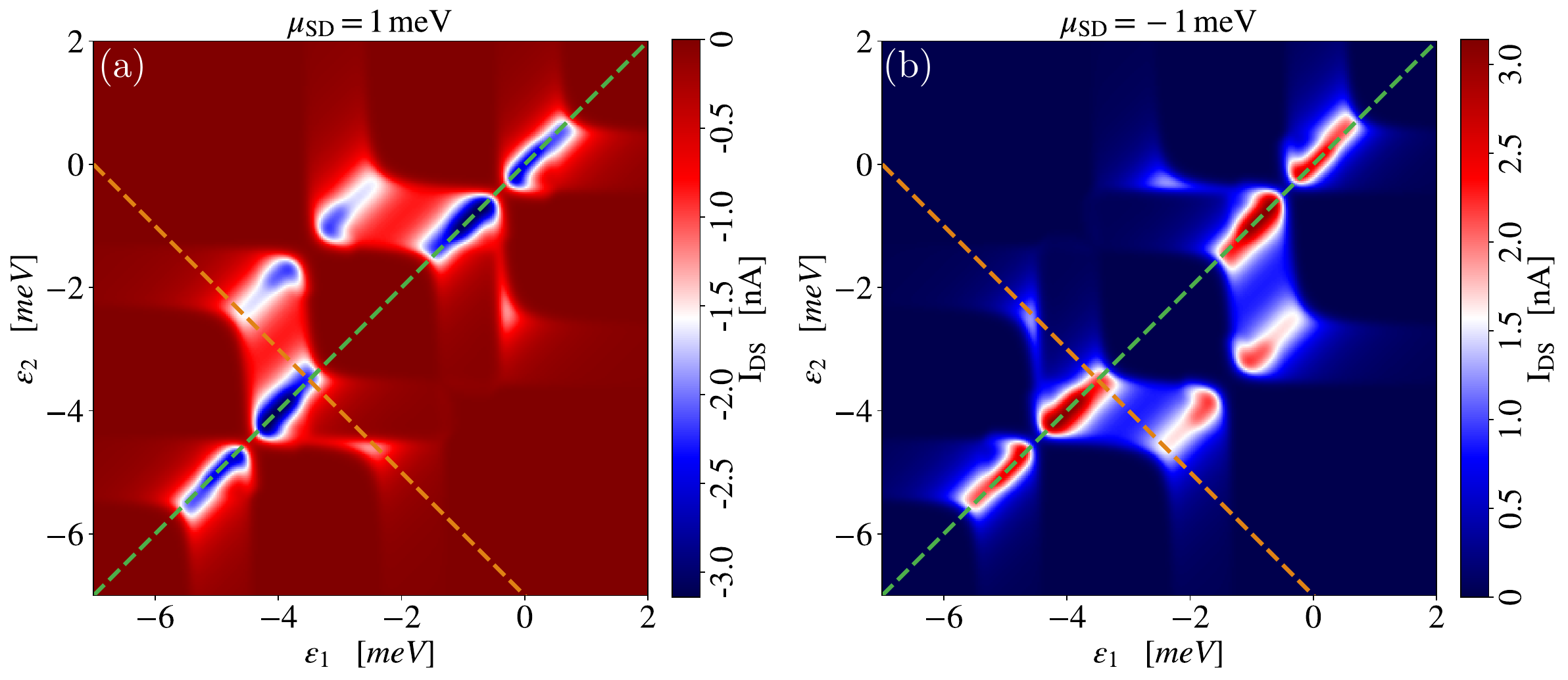}
    \caption{DISET current response for onsite potential sweep. The cutlines A and B are shown again for reference. The parameters are the same as in Fig.\,\ref{fig:diset_4_charge_stability_vdsSweeps} bar $\mu_{SD}$. Note the current is blocked in certain regions depending on the sign of the bias - see cutline B in both figures. (a) $\mu_{\text{SD}} = 1\,\text{meV}$ (b) $\mu_{\text{SD}} = -1\,\text{meV}$.}
    \label{fig:diset_4_charge_stability_PMVds}
\end{figure}

The cutline A in Fig.\,\ref{fig:diset_4_charge_stability_vdsSweeps} produces a Coulomb diamond response in a similar manner to the SISET but there are many more transitions visible within the diamonds themselves due to the 4 electron interacting dot region. However, now that the DISET has two onsite potentials, we can explore detuning cutlines that were not possible in the SISET case. For example, see cutline B, which we plot in Fig.\,\ref{fig:diset_4_charge_stability_vdsSweeps}. We see certain transitions disappear for large $\mu_{\text{SD}}$. To see how this manifests in the charge stability diagram via current for these large bias detunings, we plot for $\mu_{\text{SD}} = \pm1\,\text{meV}$ in Fig.\,\ref{fig:diset_4_charge_stability_PMVds}. The same cutlines are shown in Fig.\,\ref{fig:diset_4_charge_stability_PMVds} for reference. Note the widening of the charge transitions and the increase in the current magnitude due to the larger bias window.

Finally, we consider a double quantum dot with the drain reservoir disconnected, effectively sending $\gamma_R \rightarrow 0$. Therefore, all electron transitions will take place via the source reservoir.

\subsection{Double-Island Single-Electron Box}
\begin{figure}[!htb]
    \centering
    \includegraphics[width=1.0\linewidth]{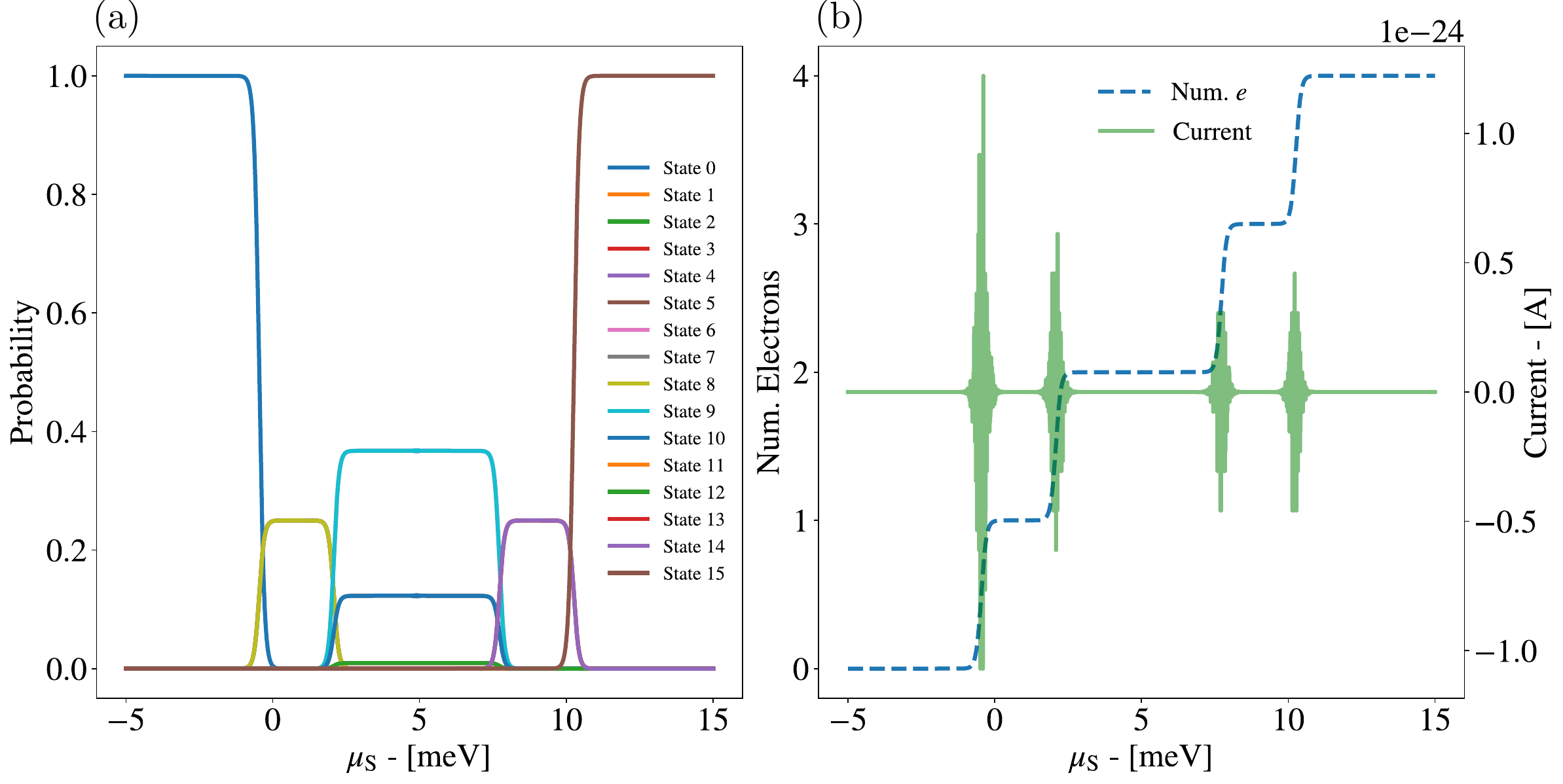}
    \caption{Parameters: $\epsilon_1=\epsilon_2=0, U_1=U_2=3.1\,\text{meV}, U_{12} = 0.9\,\text{meV}, t = 200\,\mu\text{eV}, \tau = 10\,\mu\text{eV}, \gamma_R = 0, \text{B} = 1\mu\,\text{T}, \text{T}=500\,\text{mK}$. (a) Variation in all 16 populations for the 4 electron double quantum dot system. (b) Current spikes coincide with steps in the expected number of electrons in the double dot system over $\mu_{S}$ sweep. }
    \label{fig:diseb_chem_states_current_loading}
\end{figure}

\begin{figure}[!htb]
    \centering
    \includegraphics[width=0.9\linewidth]{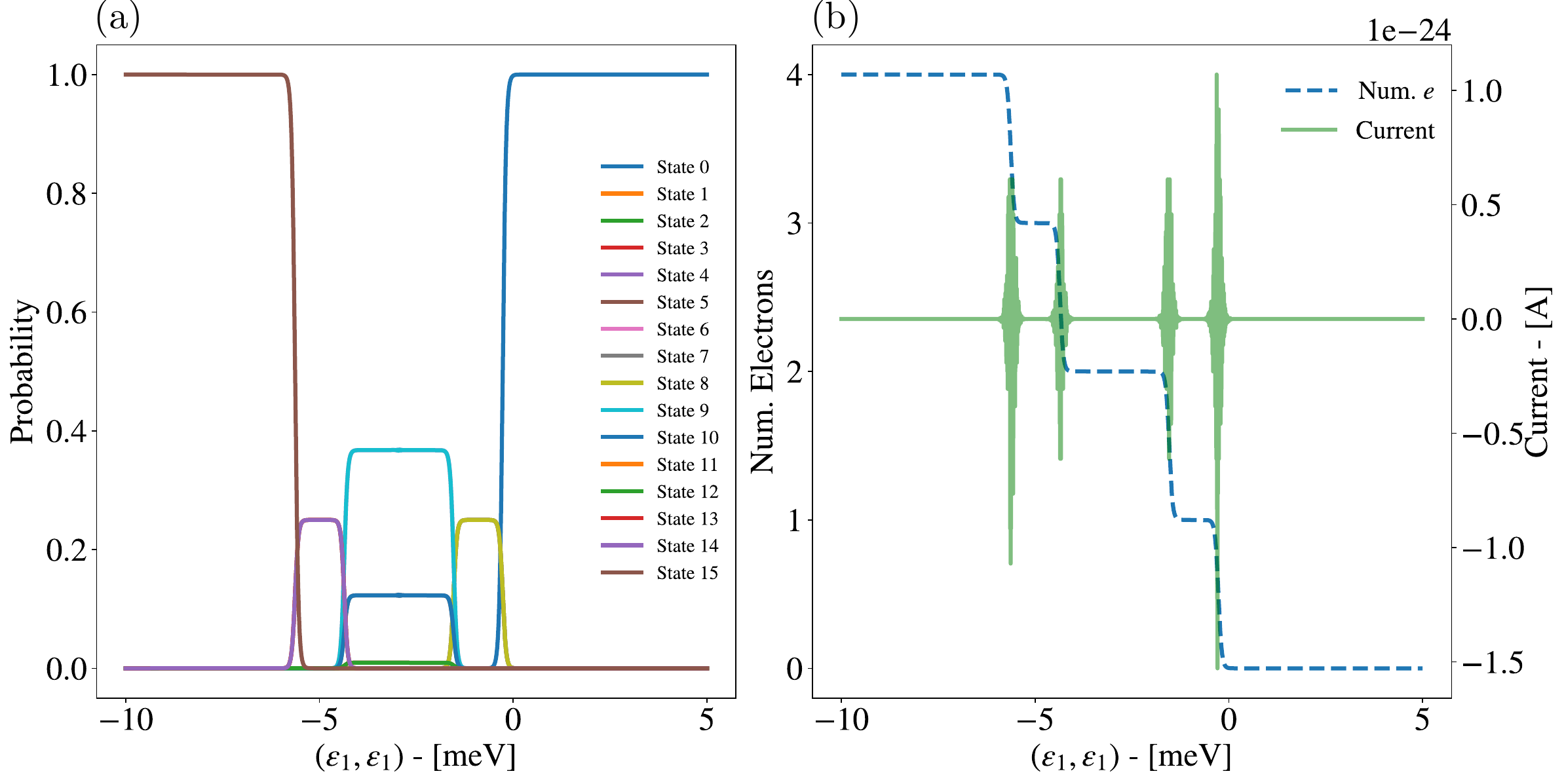}
    \caption{Same parameters as Fig.\,\ref{fig:diseb_chem_states_current_loading} with $\mu_{S} = 0$. (a) Population variation over onsite sweep along cutline A in Fig.\,\ref{fig:diset_4_charge_stability_vdsSweeps}. (b) Electrons now leave the double dot system with an increase in onsite potential, in constrast with Fig.\,\ref{fig:diseb_chem_states_current_loading}(b).}
    \label{fig:diseb_onsite_states_current_loading}
\end{figure}

In a similar manner to the SISEB, we investigate detuning cutlines for the DISEB as current cannot flow through the double dot region in box mode. There are still 16 populations, as for the DISET, visible in Fig.\,\ref{fig:diseb_chem_states_current_loading}(a) and Fig.\,\ref{fig:diseb_onsite_states_current_loading}(b), for chemical potential and onsite potential sweeps respectively.  We are now loading up to 4 electrons in the DQD, compared with the SISEB.

Spikes in the current response are visible for both Fig.\,\ref{fig:diseb_chem_states_current_loading}(b) and Fig.\,\ref{fig:diseb_onsite_states_current_loading}(b) as electrons are loaded and removed respectively. Note that we are not stating in these figures which charge state the system is in as an electron is loaded or removed. While we do plot the charge populations, note that the system is in a superposition of charge states at times over the detuning sweep. For example, see the centre of Figs.\,\ref{fig:diseb_chem_states_current_loading}(a) and \ref{fig:diseb_onsite_states_current_loading}(a), where the superposition over a number of the charge states is clearer.

\section{Linear Response}\label{app:linearresponse}

\begin{figure}[!htb]
    \centering
    \includegraphics[width=0.8\linewidth]{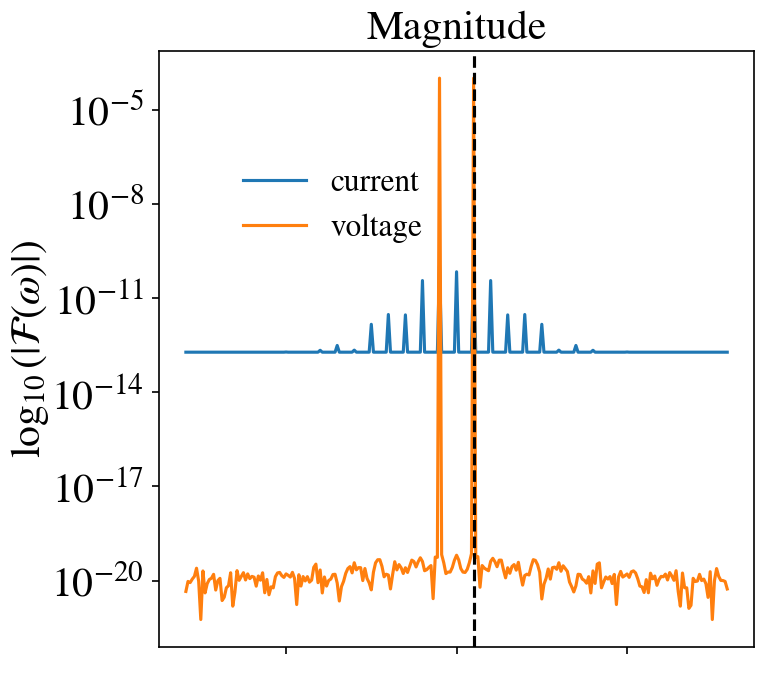}
    \caption{With a probe amplitude of 100\,$\mu$\text{V} the small perturbation assumption is broken. There are a significant number of overtones visible in the current response indicating the device is driven outside its linear response regime.}
    \label{fig:Overamplitude}
\end{figure}

We found in simulation that 1\,$\mu$V produced results consistent with experiment and is also consistent with values seen in literature, as discussed in the main text. As an illustration of a probe amplitude that breaks the small perturbation assumption, we show in Fig.\,\ref{fig:Overamplitude} the response for a probe amplitude of 100\,$\mu$V. Note the number of overtones present in the Fourier transform of the current signal and their relative sizes. This demonstrates the device is no longer being driven within a linear response regime and so it is not possible to associate lumped circuit components with it as we did in the main text. Compare the Fourier transform here with that in Fig.\,\ref{fig:QuantumMeasurement_ReflectoFullFlow}.

\end{document}